\documentclass[12pt, a4paper]{article}

\usepackage[numbers]{natbib}
\usepackage{rotating}
\usepackage{graphicx}

\graphicspath{{Figures/}}
\usepackage{lineno}

\usepackage{ifthen} 
\newboolean{pdflatex}
\setboolean{pdflatex}{true} 

\newboolean{articletitles}
\setboolean{articletitles}{true} 

\newboolean{uprightparticles}
\setboolean{uprightparticles}{false} 

\newboolean{inbibliography}
\setboolean{inbibliography}{false} 

\usepackage{mciteplus}

\def\pt{p_{\rm T}}
\def\ptmuj{\pt(\mu)/\pt(j_\mu)}

\begin{document}

\title{Forward EW Physics at the LHC}


\maketitle
\begin{center}
Stephen Farry$^{1}${\small \\On behalf of the ATLAS, CMS and LHCb collaborations \\ $^{1}$Department of Physics, University of Liverpool, L69 7ZE, United Kingdom.}
\end{center}

\begin{abstract}
Measurements of electroweak production in the forward region at the LHC provide unique and complementary information to those performed in the central region. Studies have been performed not just by LHCb, a dedicated forward detector, but also by ATLAS and CMS, which are primarily situated in the central region but can exploit forward calorimetry coverage to contribute to the understanding of SM processes in the forward region.
\end{abstract}

\section{Introduction}

\begin{figure}
\includegraphics[width=0.43\textwidth]{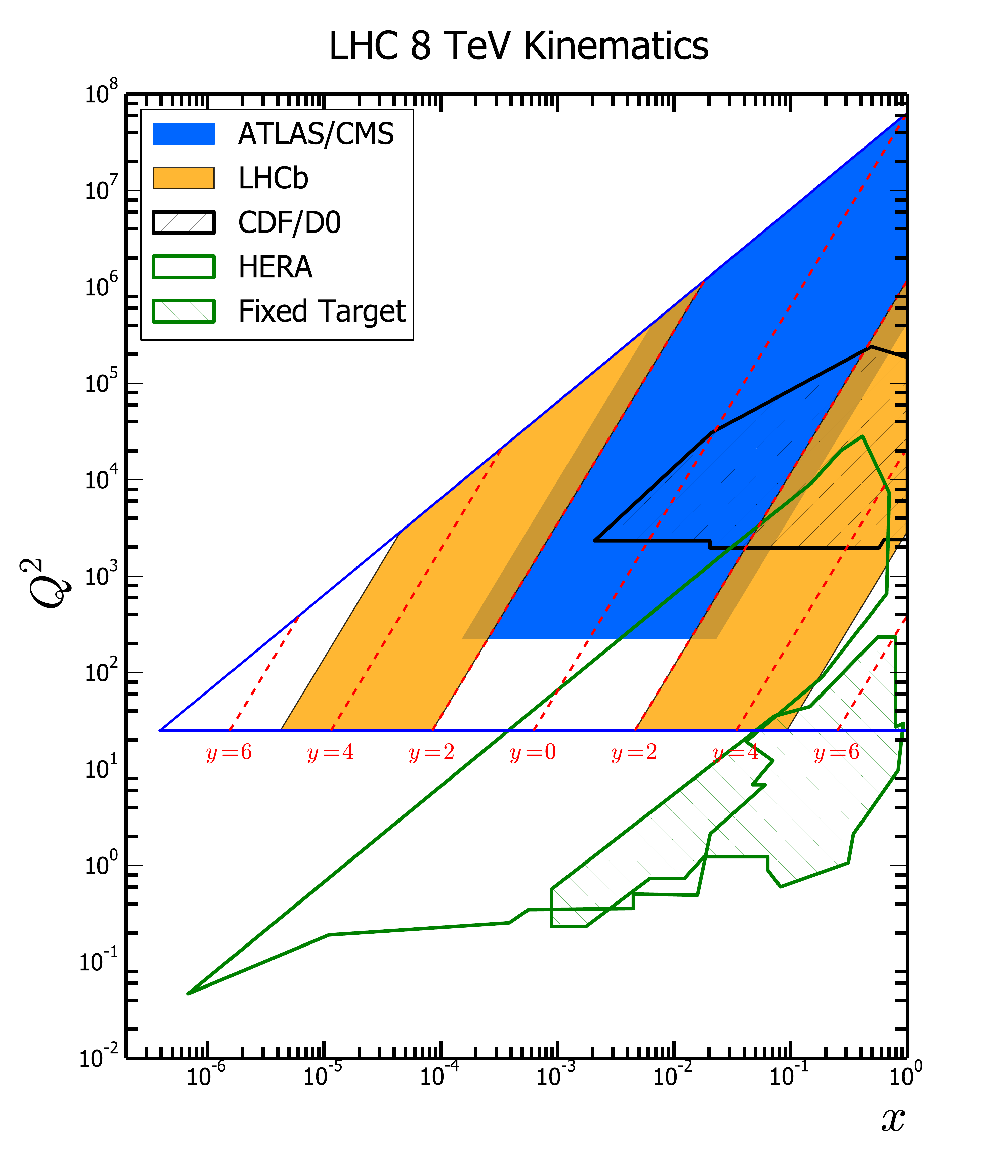}
\includegraphics[width=0.57\textwidth]{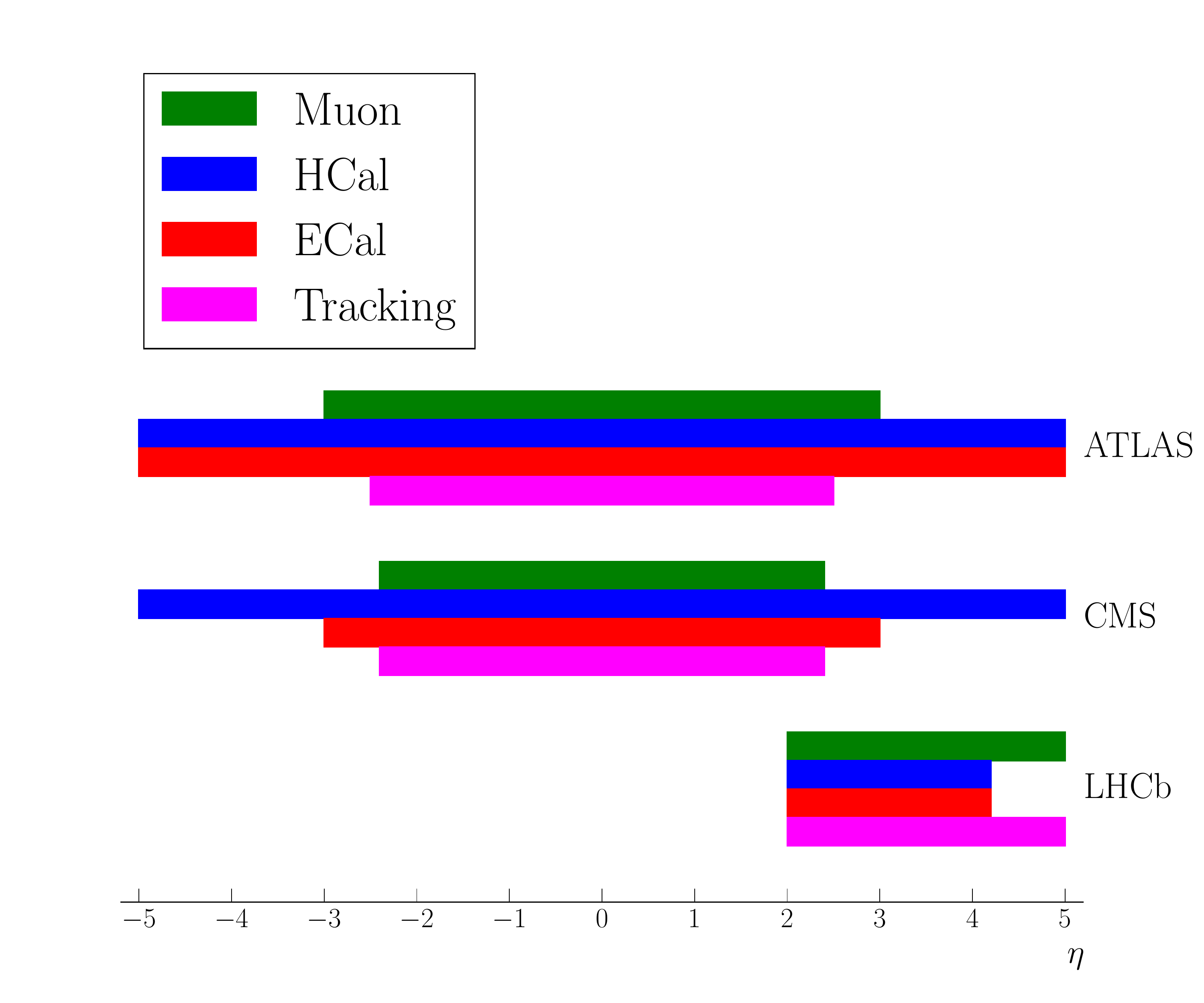}
\caption{ (Left) The region of the ($x$, $Q^2$) space probed by previous experiments as well as ATLAS, CMS and LHCb at a centre-of-mass energy of 8 TeV. (Right) The approximate rapidity coverage of the different components of the ATLAS, CMS and LHCb detectors.}
\label{fig:rapidity}
\end{figure}

Studies of the production and decay of electroweak bosons in the forward region provide both complementary and unique information to those performed in the central region. The measurements can be used to provide constraints on the parton distribution functions (PDFs) as they probe a distinct region of ($x$, $Q^2$) phase space, where $x$ is the longitudinal momentum fraction of the proton carried by the proton, and $Q^2$ is the energy scale of the interaction. As forward bosons are produced through the annihilation of two partons with asymmetric longitudinal proton momentum fractions, studies of their production simultaneously probes the low- and high-$x$ regions of the ($x$, $Q^2$) plane. This is illustrated in the left diagram of Figure~\ref{fig:rapidity}, where the different regions explored by the LHC experiments at a centre-of-mass energy of 8~TeV are shown.

The extension of measurements into the forward region is also particularly relevant for the electroweak production of $W$ and $Z$ bosons. The signature for these process is the production of the boson in association with two jets with a large rapidity gap arising from colour flow considerations. The inclusion of forward jets in the measurements allows for a greater phase space with which to identify these rapidity gaps.

At a proton-proton collider such as the LHC, the forward region is also a more sensitive probe of the forward-backward asymmetry in $Z$ decays. $Z$ bosons produced in the forward region are more likely to follow the initial quark direction and consequently suffer less from the dilution of the parton level asymmetry observed in the central region. This is discussed in more detail later in the text.

The approximate rapidity coverage of the ATLAS~\cite{Aad:2008zzm}, CMS~\cite{Chatrchyan:2008aa} and LHCb~\cite{Alves:2008zz} experiments is shown in the right diagram of Figure~\ref{fig:rapidity}. LHCb is an experiment optimised for the study of $\mathcal{CP}$ violation in heavy flavour decays, and consequently is fully instrumented in the forward region. This includes tracking coverage in the pseudorapidity range between 2.0 and 5.0, as well as electromagnetic and hadronic calorimetry up to approximately 4.5. The ATLAS and CMS detectors are instrumented in the central region, and consequently their tracking coverage is limited to $|\eta| < 2.4$. However, both experiments have calorimetry coverage further forward which allows them to identify jets and electrons in the forward region. In particular, the ATLAS forward calorimeters provide both hadronic and electromagnetic calorimeter coverage up to $\eta=4.9$, while the CMS electromagnetic calorimeter coverage extends up to $\eta=3.0$ with hadronic calorimetry coverage up to $\eta=5.0$.

Measurements of inclusive $W$ and $Z$ production at LHCb are discussed first, followed by measurements of $W$ and $Z$ boson production in association with inclusive jets and $W$ production in association with heavy flavour jets. Finally the electroweak production of $W$ and $Z$ bosons is presented followed by measurements of the forward-backward asymmetry in $Z\to\ell\ell$ events.

\section{Inclusive $W$ and $Z$ Production}
\label{sec:inclwz}
\begin{figure}
\begin{center}
\includegraphics[width=0.6\textwidth]{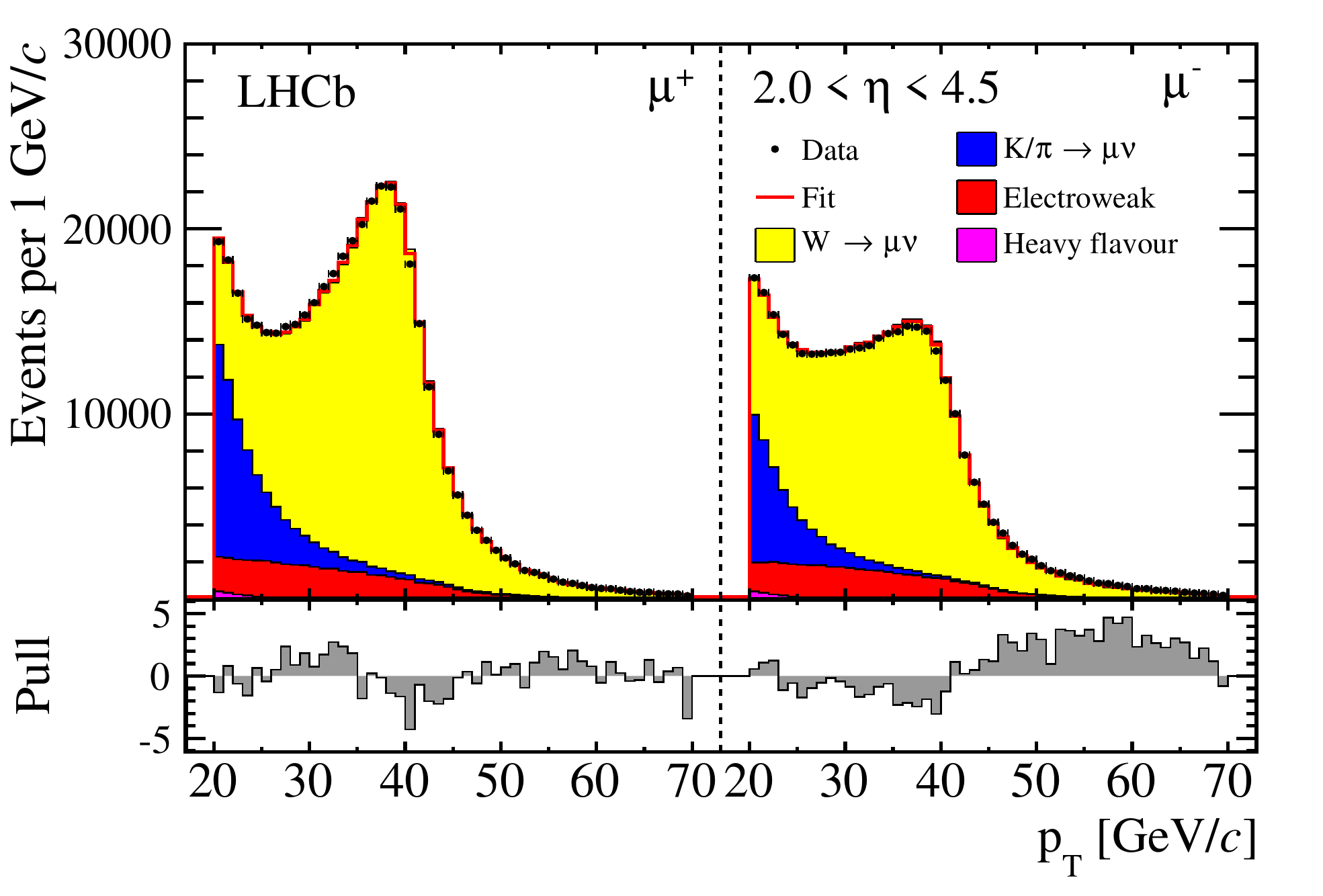}
\end{center}
\caption{Fit to the muon $p_{\rm T}$ spectrum for $W^+$ (left) and $W^-$ (right) candidates for purity extraction~\cite{Aaij:2014wba}.}
\label{fig:wfit}
\end{figure}
\begin{figure}
\includegraphics[width=0.5\textwidth]{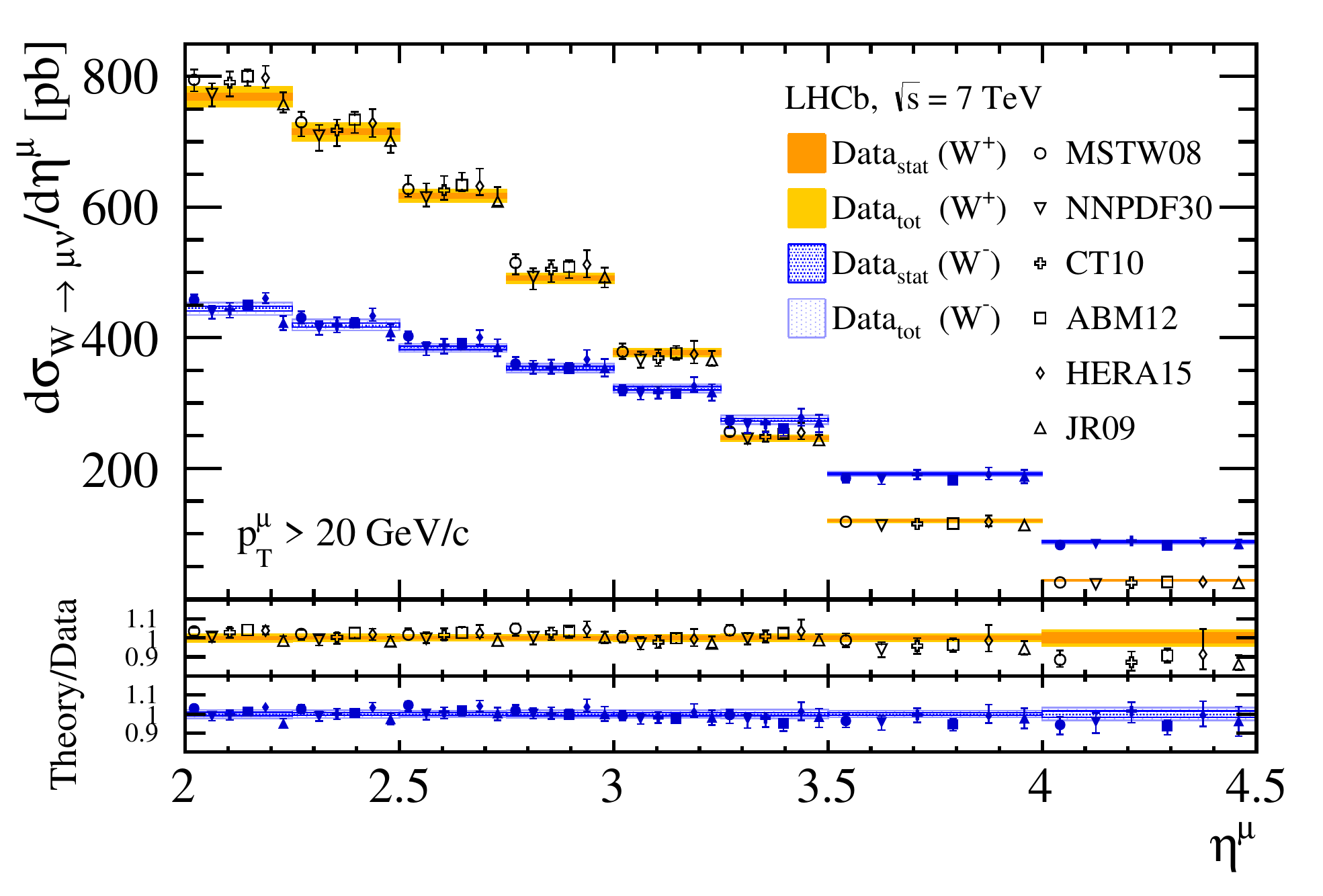}
\includegraphics[width=0.5\textwidth]{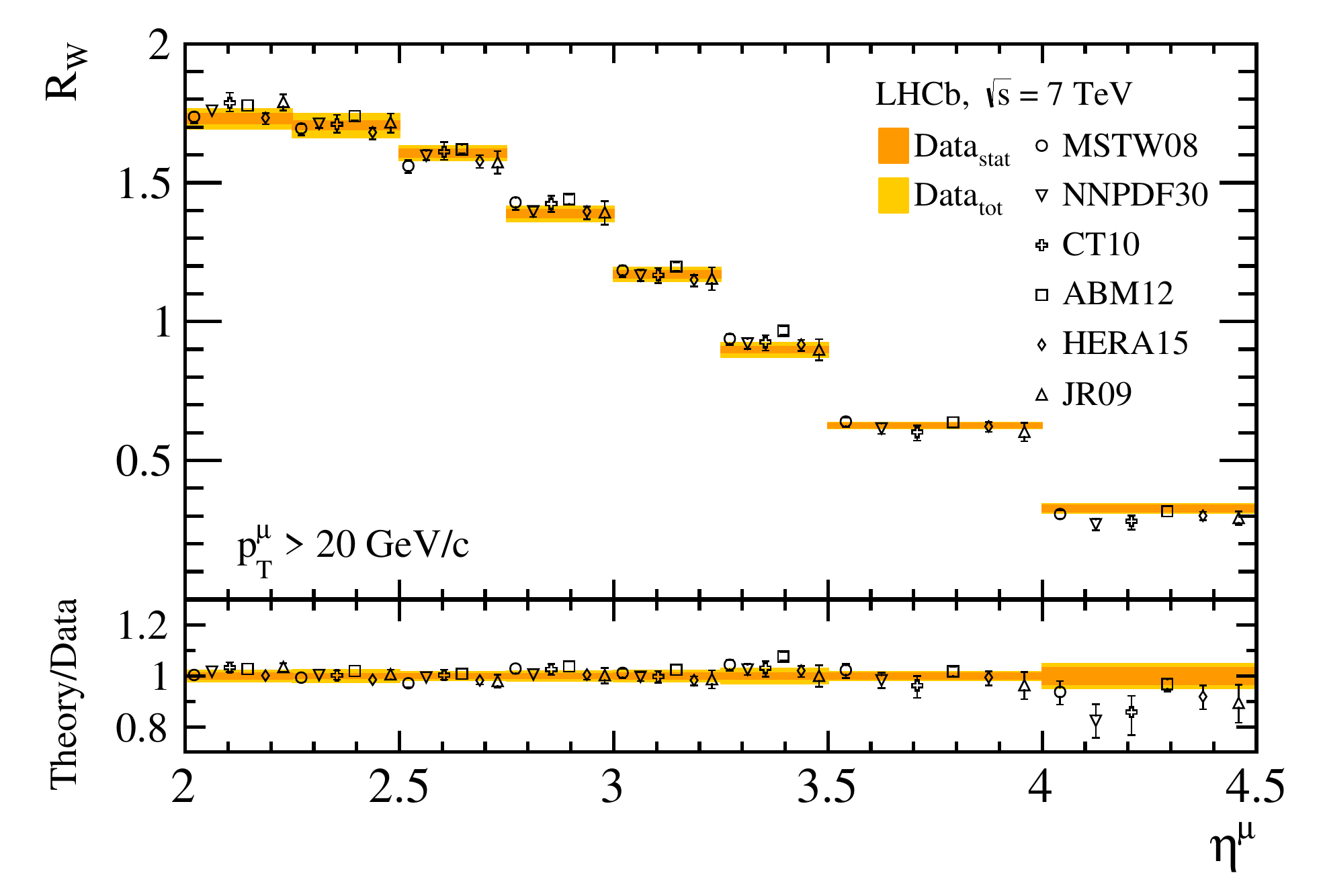}
\caption{Measurements of (left) the $W$ boson production cross-section and (right) the lepton charge asymmetry at LHCb, where both are shown as a function of lepton pseudorapidity and compared to NNLO QCD predictions using a variety of PDF sets~\cite{Aaij:2015gna}.}
\label{fig:wresults}
\end{figure}
\begin{figure}
\includegraphics[width=0.486\textwidth]{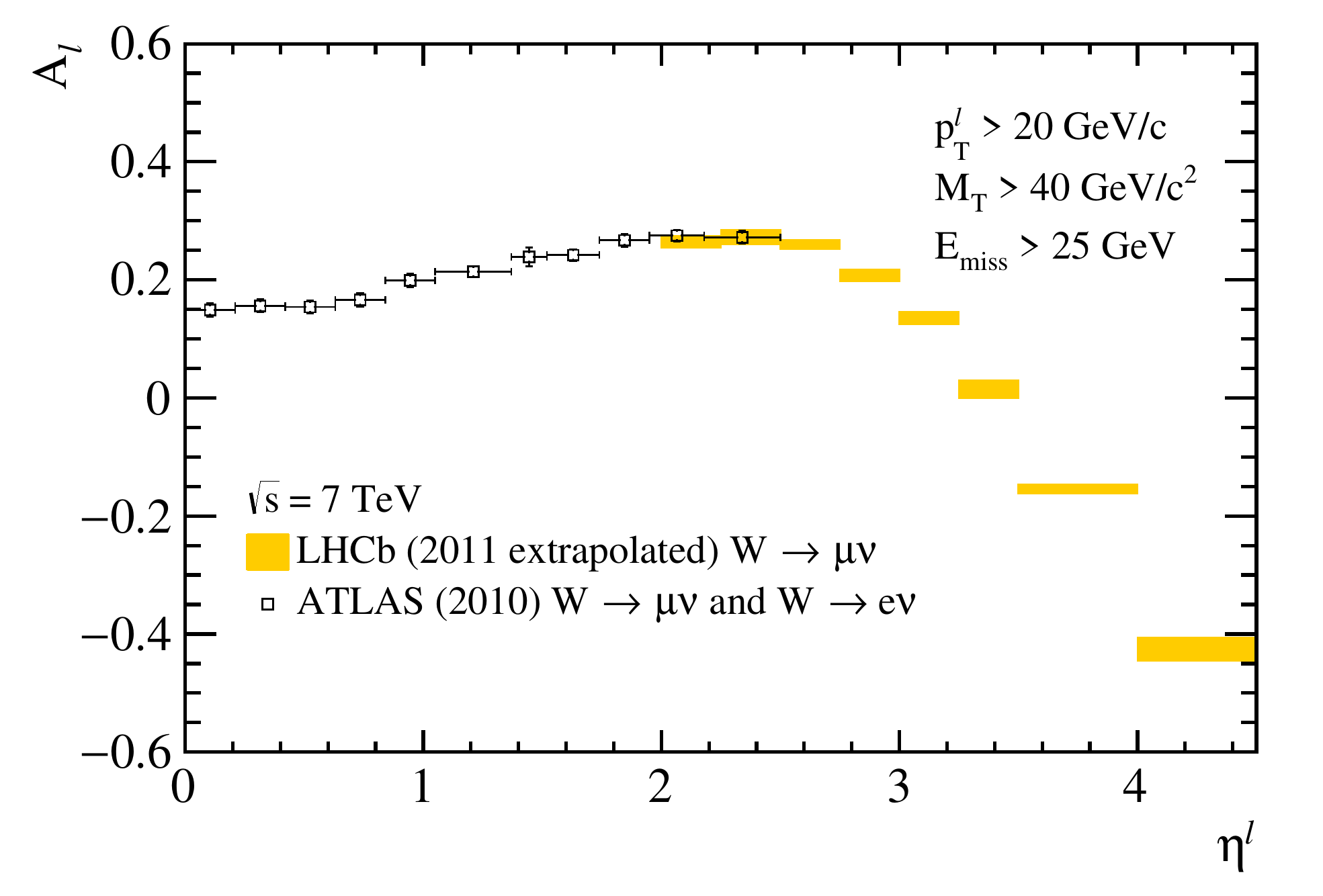}
\includegraphics[width=0.514\textwidth]{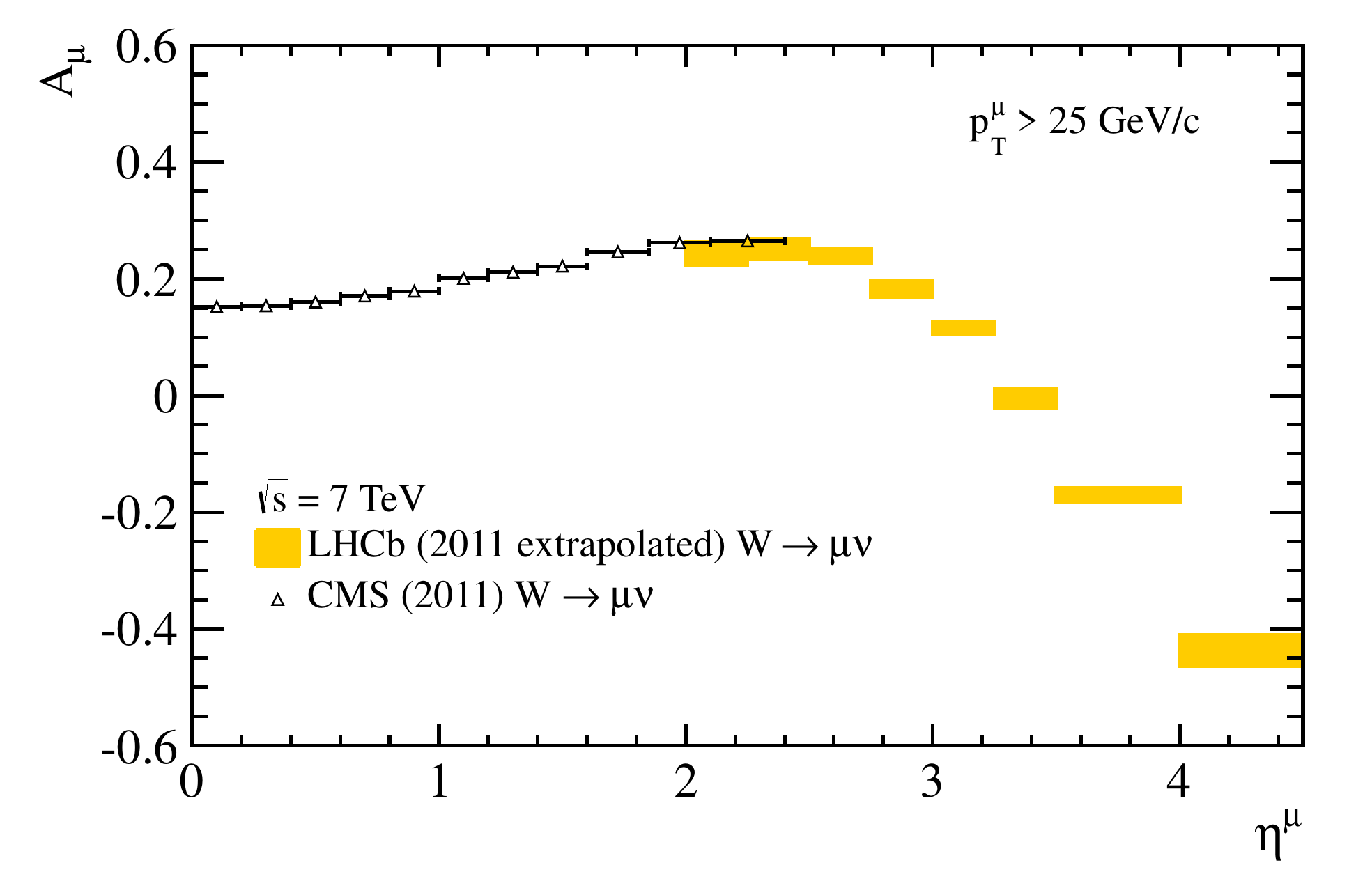}
\caption{Comparisons of the measured inclusive $W$ lepton charge asymmetry between (left) LHCb and ATLAS, and (right) LHCb and CMS. In both cases the LHCb measurement is extrapolated to match the ATLAS/CMS fiducial volume using theoretical predictions~\cite{Aaij:2015gna}.}
\label{fig:wcomparisons}
\end{figure}
\begin{figure}
\includegraphics[width=0.57\textwidth]{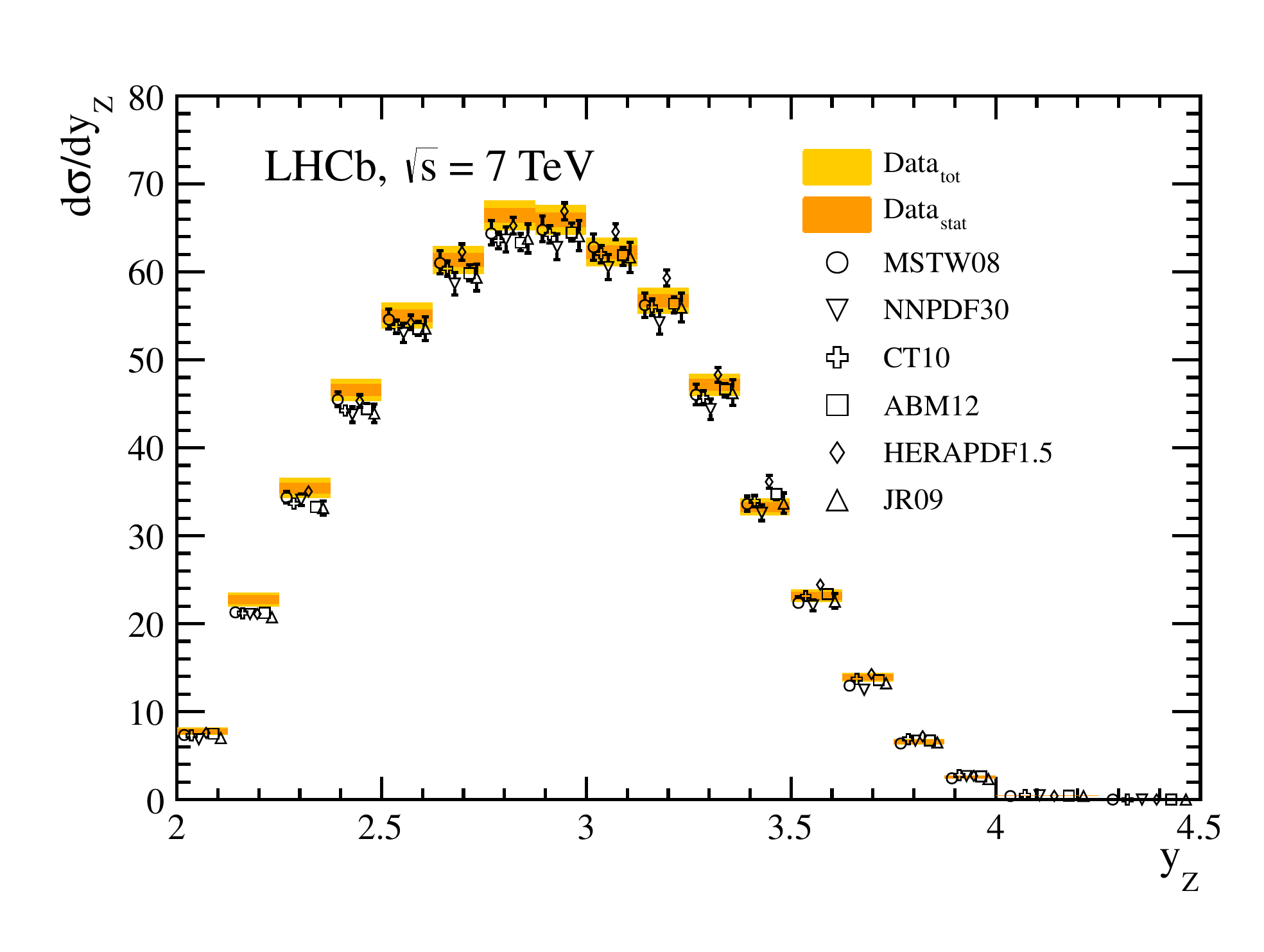}
\includegraphics[width=0.43\textwidth]{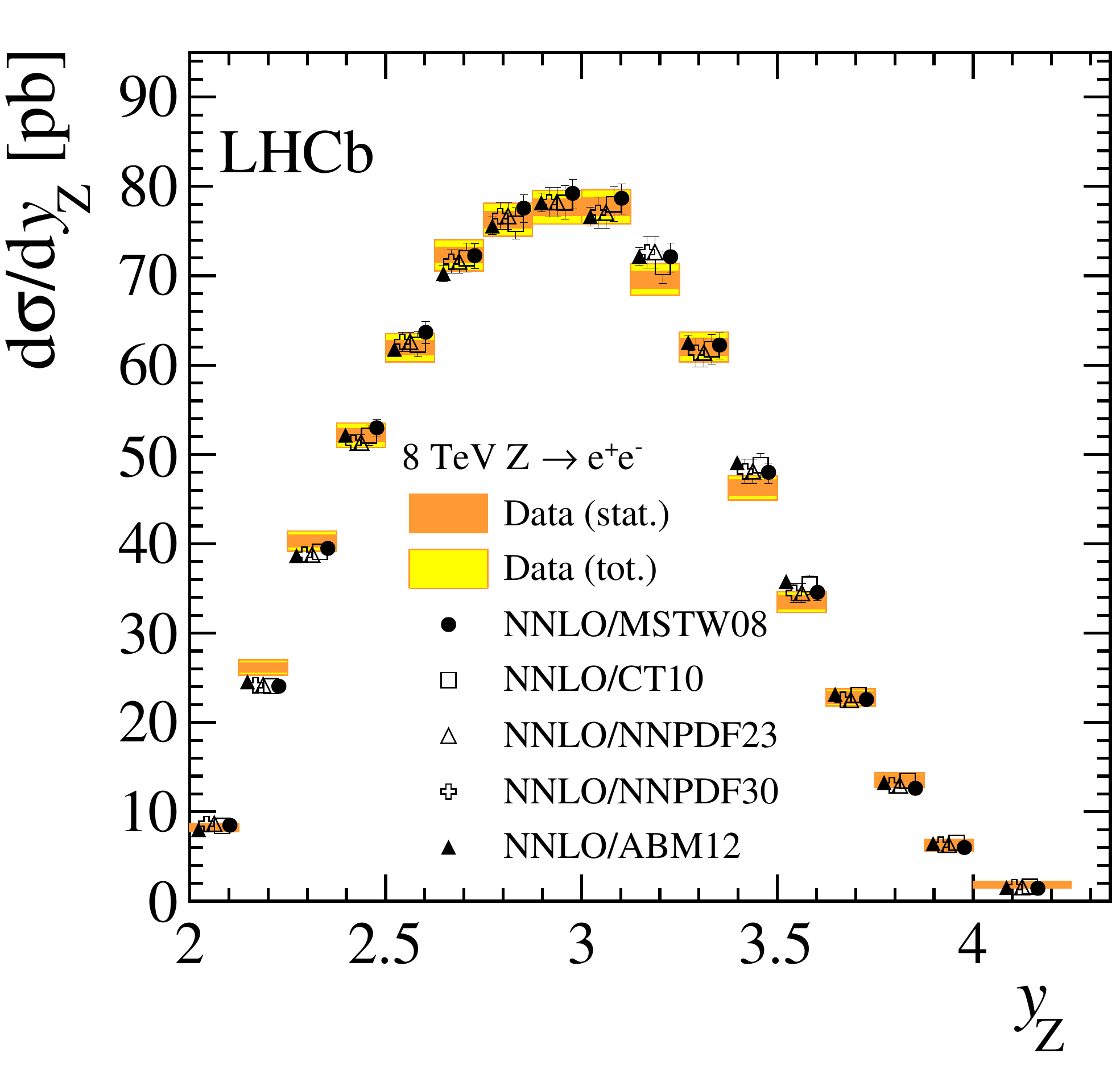}
\caption{Measurements of the $Z$ boson production cross-section measured at LHCb for the (left) muon and (right) electron decay mode. Comparisons are made to NNLO QCD predictions with a range of PDF sets~\cite{Aaij:2015gna,Aaij:2015vua}.}
\label{fig:zresults}
\end{figure}

Measurements of the inclusive production of $W$ and $Z$ bosons in the forward region have recently been performed by the LHCb experiment at $\sqrt{s} = 7$~TeV in the muon channel~\cite{Aaij:2014wba,Aaij:2015gna} and of $Z$ production at both $\sqrt{s}=7$ and 8~TeV in the electron channel~\cite{Aaij:2012mda, Aaij:2015vua}. The measurement of $W$ production is performed by selecting muons which have a transverse momentum of greater than 20~GeV and a pseudorapidity in the region $2.0<\eta<4.5$. The purity of the $W$ boson sample is determined by performing a template fit to the $p_{\rm T}$ of the selected events using signal and background shapes obtained from both data and simulation. The signal shape and the dominant background, arising from muons produced through the decay-in-flight of pions and kaons, are allowed to float in the fit, with the other backgrounds normalised using data-driven techniques. The fit is shown for both positive and negative muons in Figure~\ref{fig:wfit} with a purity of approximately 77\% achieved for both $W^+$ and $W^-$. The extracted yields are corrected for detector reconstruction and selection efficiency, as well as final state radiation in order to facilitate a comparison with fixed order QCD predictions. The differential cross-sections and lepton charge asymmetry are shown in Figure~\ref{fig:wresults} as a function of lepton pseudorapidity. The results are compared to predictions obtained at NNLO in perturbative QCD using the \textsc{FEWZ}~\cite{Gavin:2010az} generator and a number of different PDF sets with a good level of agreement observed. The experimental precision is dominated by the luminosity measurement, as well as the effect of the beam energy uncertainty, with both sources contributing uncertainties of 1.16 and $\sim1\%$ respectively. The presence of a small overlap in the coverage of the ATLAS, CMS and LHCb detectors also allows a comparison to be made between the three experiments. This comparison is shown in Figure~\ref{fig:wcomparisons} for ATLAS~\cite{Aad:2011dm} and CMS~\cite{CMS:2011aa} where in both cases the LHCb data is corrected using theoretical predictions to account for differences in the chosen measurement fiducial volumes. In order to perform measurements of $Z$ boson production, the same kinematic selection is applied as in the case of $W$ production, where electrons are also selected, and a second, opposite-sign lepton is also required to be present and the pair are required to form an invariant mass in the region of the $Z$ peak. In the dimuon channel, this mass range is chosen to be between 60 and 120 GeV, while for the electron channel, incomplete bremsstrahlung recovery results in the smearing of the mass spectrum to lower masses, and so the dielectron invariant mass is just required to be larger than 40~GeV. The purity in both channels is determined using data-driven methods, and is over 99\% for the dimuon channel, and approximately 95\% in the electron channel. The measurements are presented and compared to SM predictions using a range of different PDF sets as a function of boson rapidity in Figure~\ref{fig:zresults}, with a similar level of agreement observed as in the case of $W$ boson production. Uncertainties on the overall normalisation due to the luminosity and the beam energy uncertainty are again seen to dominate. It should be noted that the beam energy is only applied to the $Z\to\mu\mu$ measurement, where a more precise measurement of the luminosity is exploited. Measurements of $W$ and $Z$ production in the muon channel at 8~TeV are not discussed here but have also since been performed by LHCb~\cite{Aaij:2015zlq}.

\section{$W$ and $Z$ Production in Association with Jets}
\begin{figure}
\begin{center}
\includegraphics[width=0.8\textwidth]{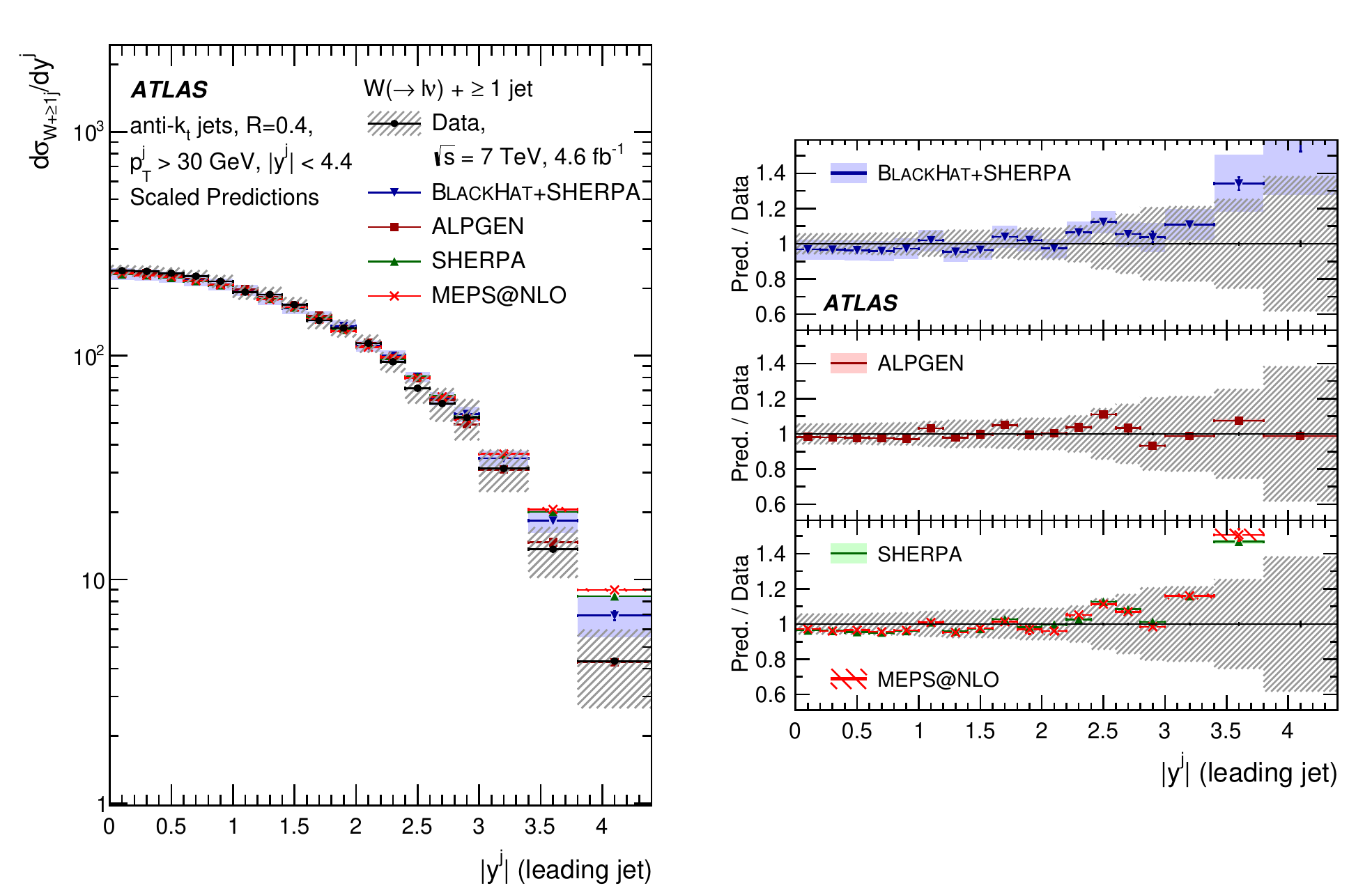}
\end{center}
\caption{Measurements of the cross-section for $Wj$ production at ATLAS shown for the rapidity of the leading jet. Deviations from the SM predictions are observed in the most forward bins~\cite{Aad:2014qxa}.}
\label{fig:atlaswj}
\end{figure}

\begin{figure}
\begin{center}
\includegraphics[width=0.8\textwidth]{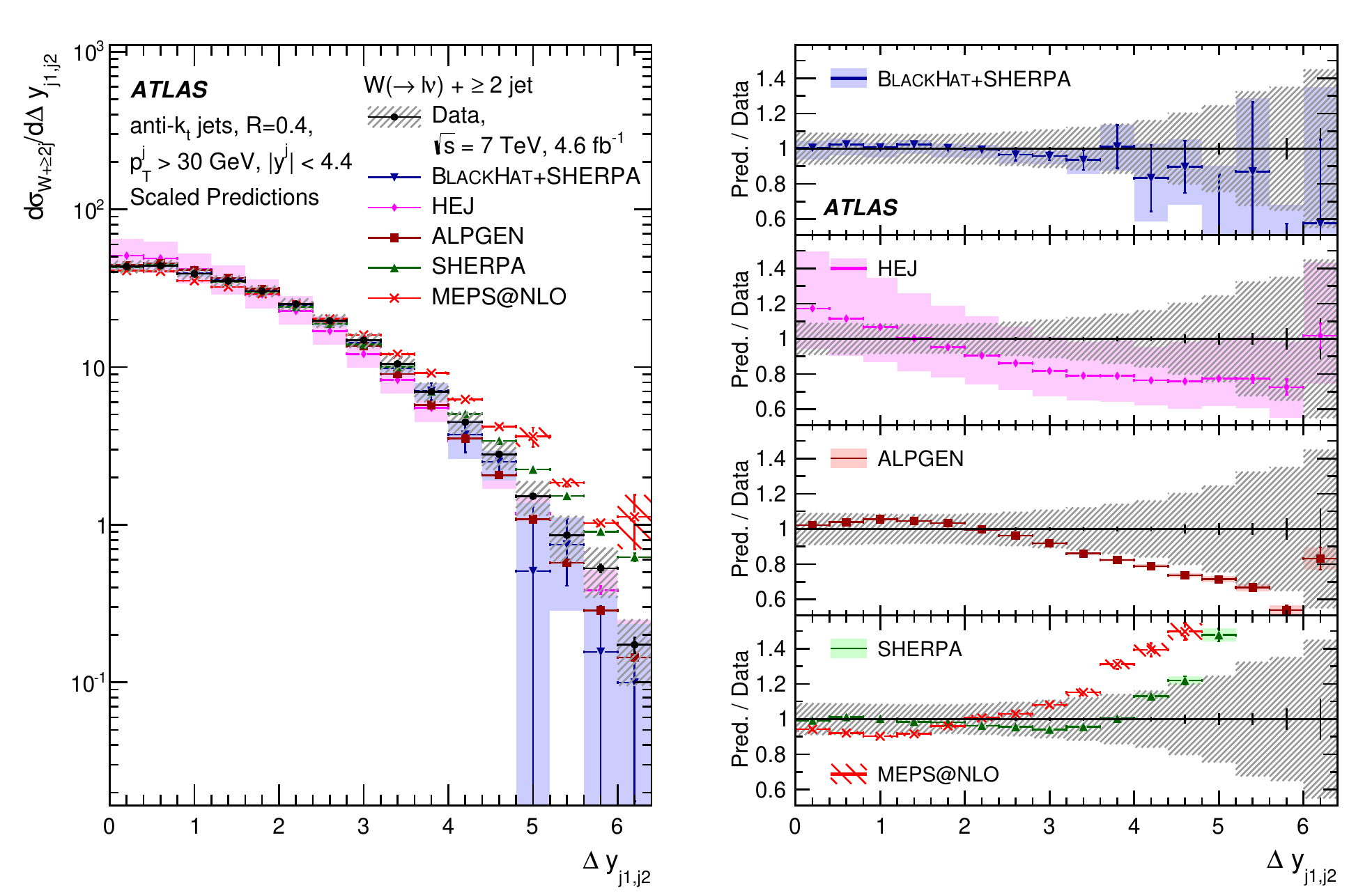}
\end{center}
\caption{Measurements of the cross-section for $Wj$ production at ATLAS shown for the  the rapidity separation of the leading and subleading jet in events with at least two jets~\cite{Aad:2014qxa}.}
\label{fig:atlaswj2}
\end{figure}

\begin{figure}
\includegraphics[width=0.5\textwidth]{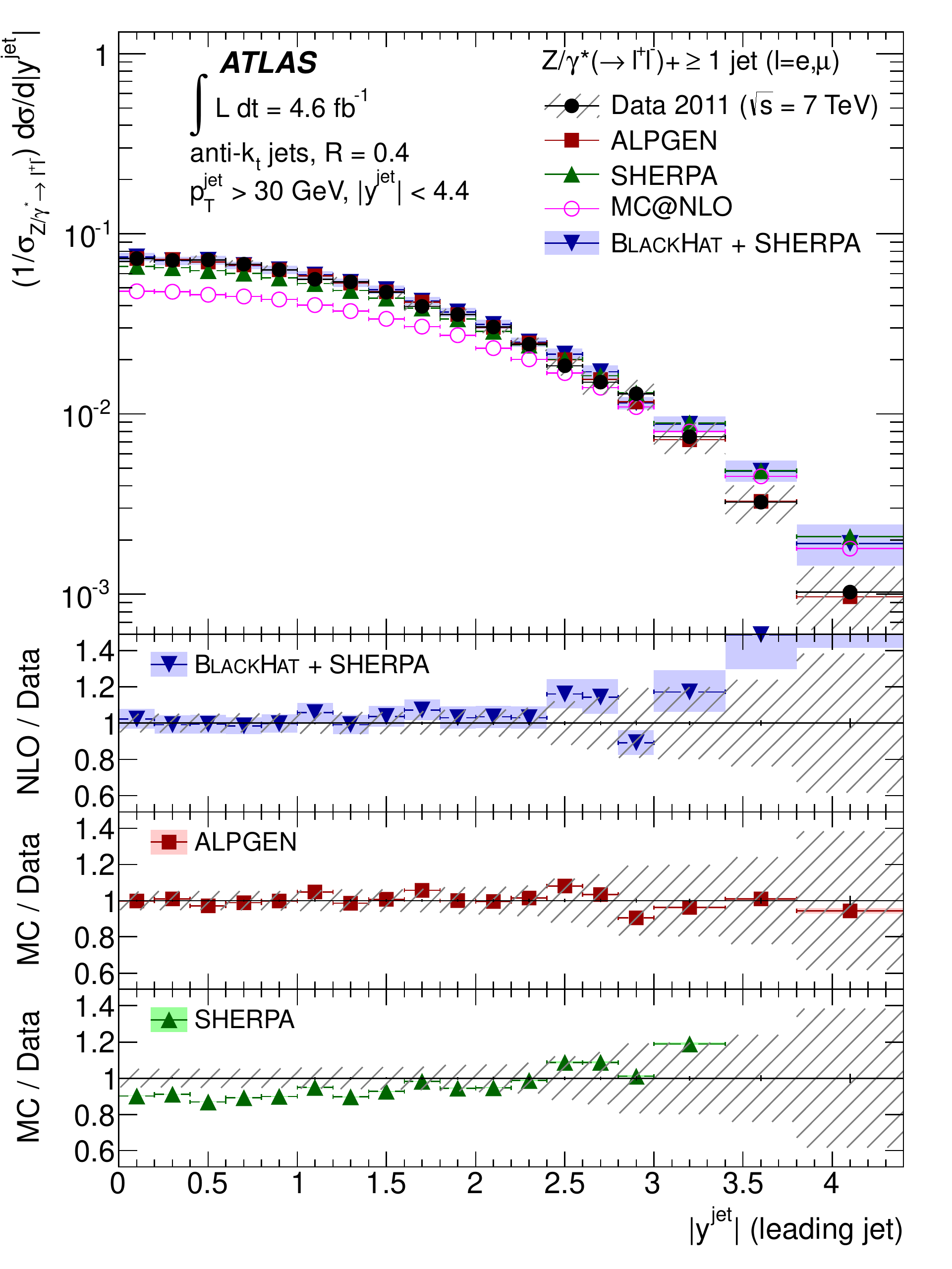}
\includegraphics[width=0.5\textwidth]{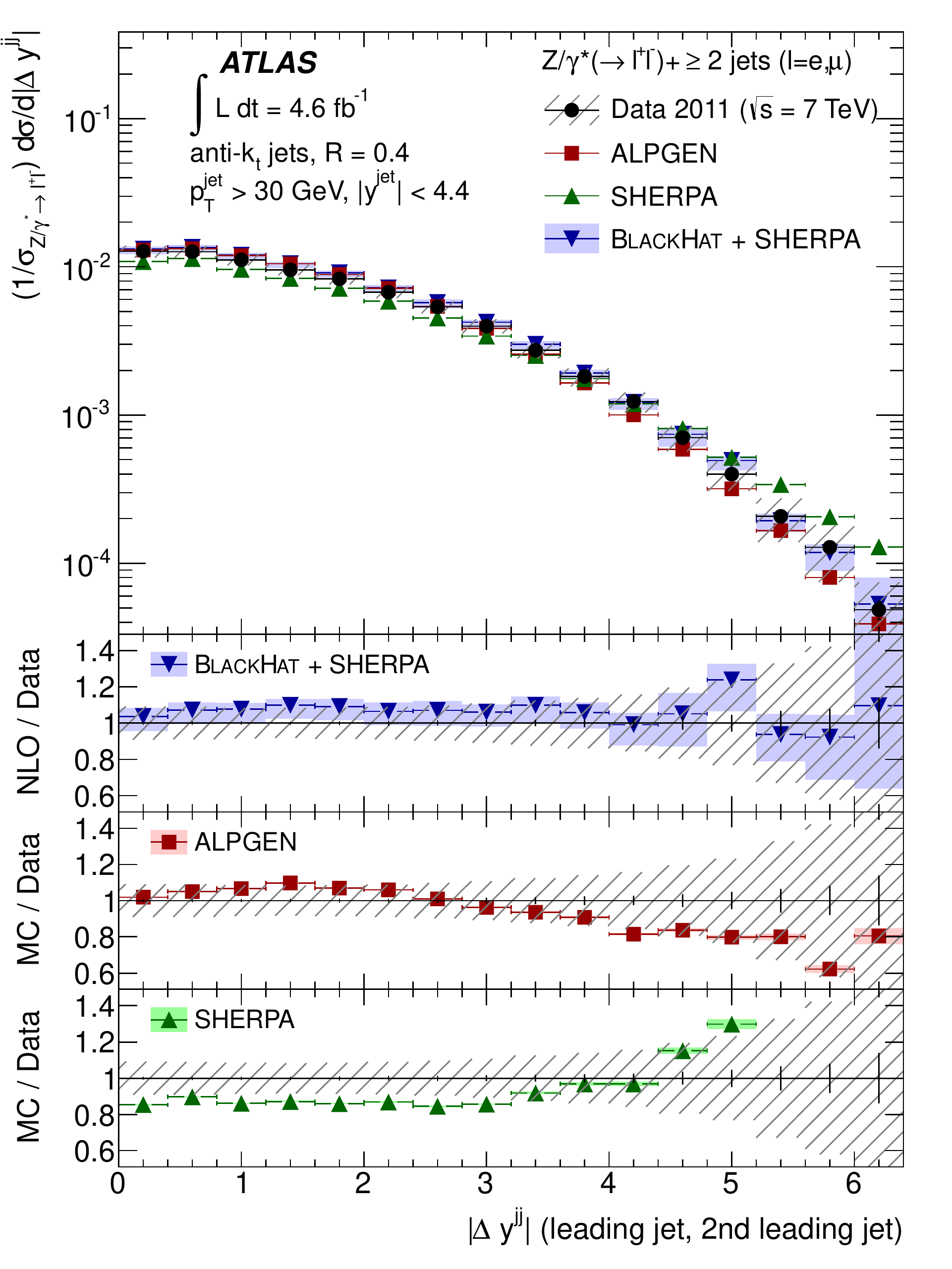}
\caption{Measurements of the cross-sections for $Zj$ production at ATLAS shown for (left) the rapidity of the leading jet and (right) the rapidity separation of the leading and subheading jet in events with at least two jets. Deviations from the SM predictions are observed in the most forward bins~\cite{Aad:2013ysa}.}
\label{fig:atlaszj}
\end{figure}

The ATLAS collaboration has performed measurements of $W$~\cite{Aad:2014qxa} and $Z$~\cite{Aad:2013ysa} production in association with jets ($Wj$, $Zj$) where the bosons are reconstructed in both the muon and electron decay modes. The muons (electrons) are required to satisfy $|\eta| < 2.4(2.47)$ and the jets extend up to forward rapidities of 4.4. The leptons are required to have a $\pt$ in excess of 25~GeV for $W$ events, and to be greater than 30~GeV for $Z$ events, where they are also required to have a dilepton mass range between 66 and 112 GeV. The $W$ candidates are selected by vetoing on additional leptons in the final state, and requiring a missing transverse momentum, $E_{\rm T}^{\rm miss}$, in excess of 25~GeV and a transverse mass, $m_{\rm T}$, of greater than 40~GeV.

The jets are reconstructed using the anti-$k_{\rm T}$ algorithm with a distance parameter of 0.4 and are required to have transverse momenta in excess of 30~GeV and to be separated from the leptons by a radius of 0.5 in $\eta-\phi$ space. The background contributions from QCD multi-jet and $t\bar{t}$ production are estimated using data-driven methods while the remaining backgrounds are taken from simulation The cross-section is measured as a function of a number of kinematic variables, and for jet multiplicities of up to seven. Of particular interest here are the measurements of the cross-sections versus jet rapidity and the rapidity separation of the two leading jets (for events with jet multiplicities of greater than one). These distributions are shown in Figure~\ref{fig:atlaswj} and Figure~\ref{fig:atlaswj2} for $Wj$ production and Figure~\ref{fig:atlaszj} for $Zj$ production. A general good level of agreement is observed although some of the predictions show a slight overestimation of the measured cross-sections at high rapidities.

\subsection{$W$ boson production in association with heavy flavour jets}
As the LHCb experiment is optimised for the selection of heavy flavour decays in the forward region, it is uniquely suited to perform heavy flavour tagging of jets in the forward region. This is achieved using a tagging algorithm which identifies heavy flavour jets through the presence of a secondary vertex with a radial separation, $\Delta R< 0.5$ between its direction of flight and that of the jet axis. Two boosted decision trees (BDTs) are trained using characteristics of both the jet and the secondary vertex to separate light jets from heavy flavour jets, and $b-$jets from $c-$jets. More details on the secondary vertex tagging algorithm can be found in Reference~\cite{Aaij:2015yqa}.

\begin{figure}
\includegraphics[width=0.5\textwidth]{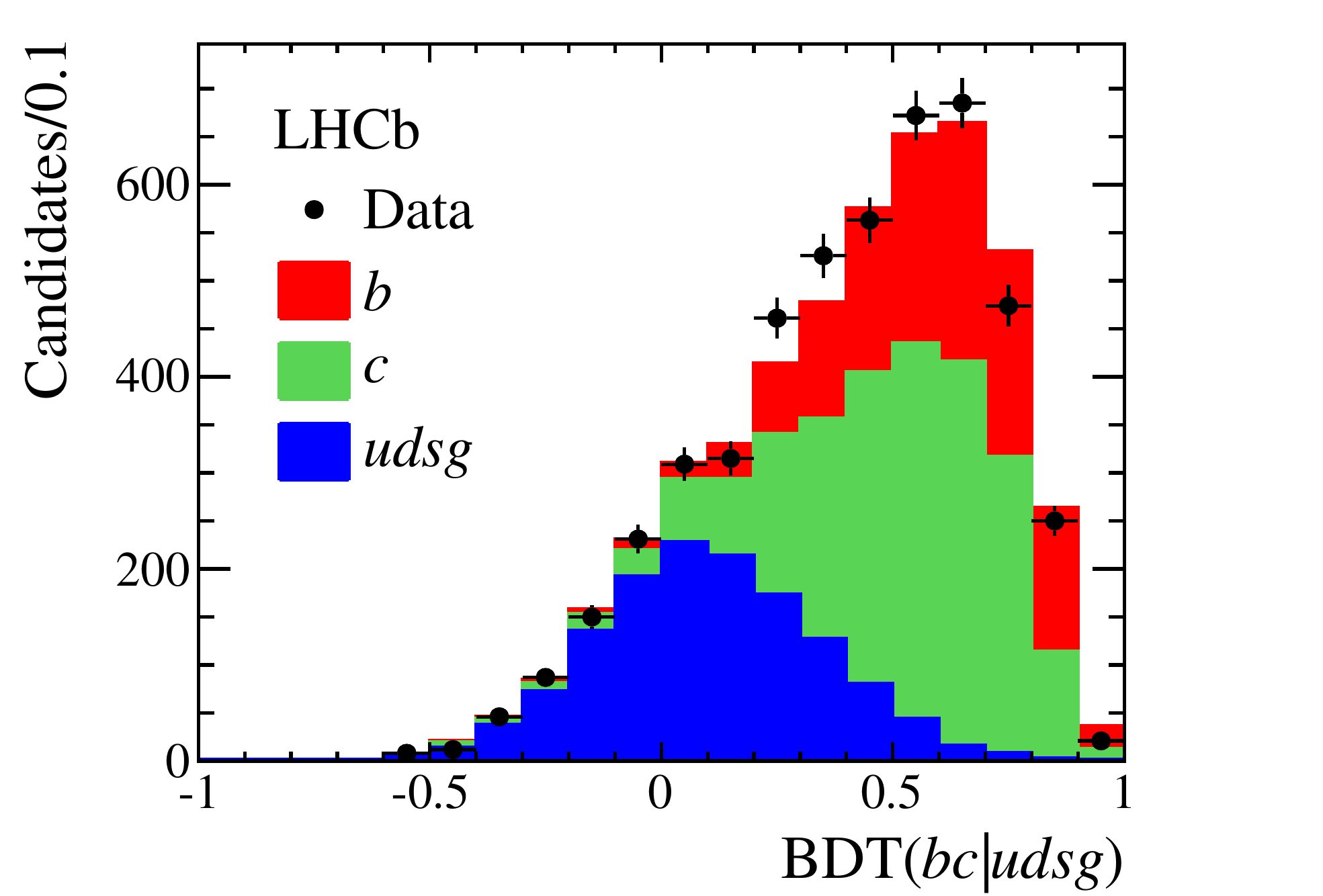}
\includegraphics[width=0.5\textwidth]{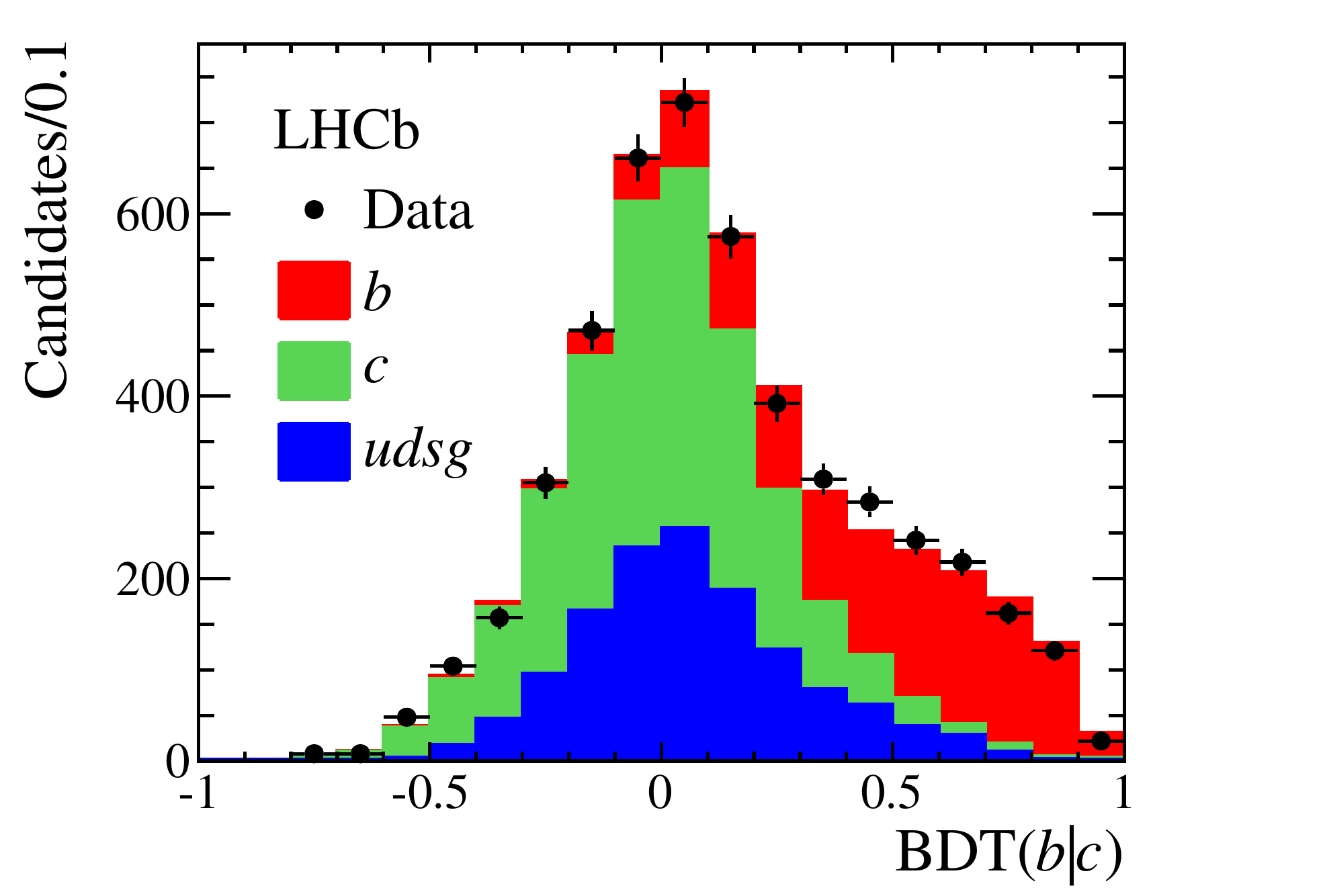}
\caption{Shown are the BDT distributions used to separate (left) light jets from heavy flavour jets and (right) $b$-jets from $c$-jets in order to measure $W$ production in association with $b-$ or $c-$jets~\cite{Aaij:2015yqa}.}
\label{fig:svtagging}
\end{figure}

These heavy flavour tagging capabilities are applied to measurements of $W$ boson production in association with $b$- and $c$-jets~\cite{Aaij:2015cha}. The $W$ boson is reconstructed using a similar selection to that described earlier, and a jet is additionally required to be present with a $p_{\rm T}$ in excess of 20~GeV and a pseudorapidity between 2.2 and 4.2. The events are also required to satisfy $p_{\rm T}(j_\mu + j) > 20$~GeV, where $j_\mu$ is a reconstructed jet containing the muon candidate. The observable is expected to be large for $W+$jet events due to the missing neutrino in the final state and consequently improves the signal purity by rejecting backgrounds arising from di-jet production where the jet momenta are balanced. 
The purity is extracted using a fit to an isolation variable defined as the ratio of $\pt (\mu)/\pt(j_\mu)$ which is representative of the isolation of the muon. The background to the sample arising from QCD backgrounds, such as the mis-identification of pions or kaons,  or the semi-leptonic decay of heavy flavour mesons is then determined using fits to this variable. The $b$ and $c$-jet yields are extracted by performing a fit to the two-dimensional BDT distributions in each bin. Additional backgrounds are considered from other electroweak processes, such as $Z\to\mu\mu$, $Z\to\tau\tau$, $W\to\tau\nu_\tau$ and top production are subtracted using data-driven techniques. Fits to the $\ptmuj$ distribution and the two-dimensional BDT distribution in the $\ptmuj > 0.9$ bin are shown in Figure~\ref{fig:svtagging}.

\begin{figure}
\includegraphics[width=0.5\textwidth]{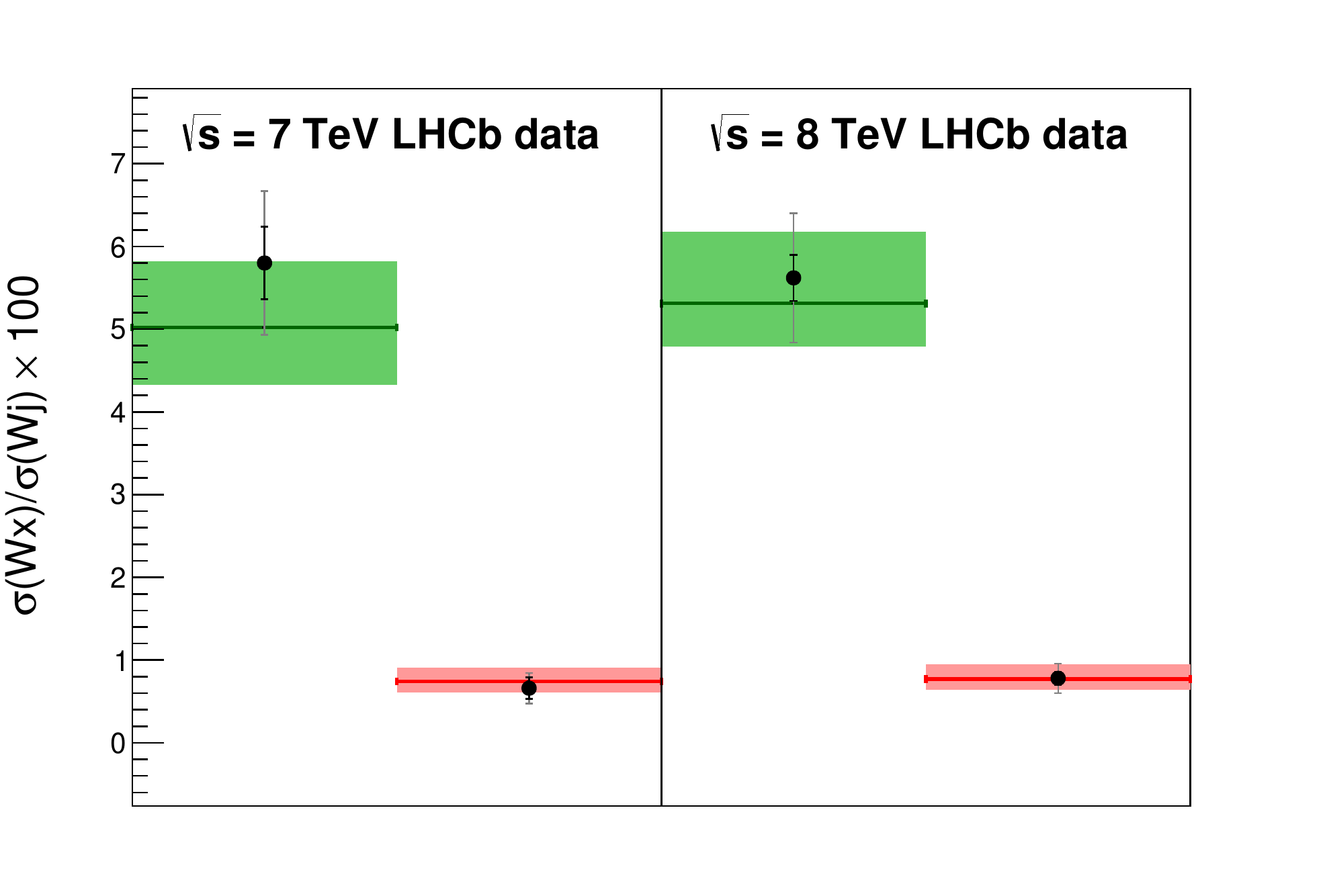}
\includegraphics[width=0.5\textwidth]{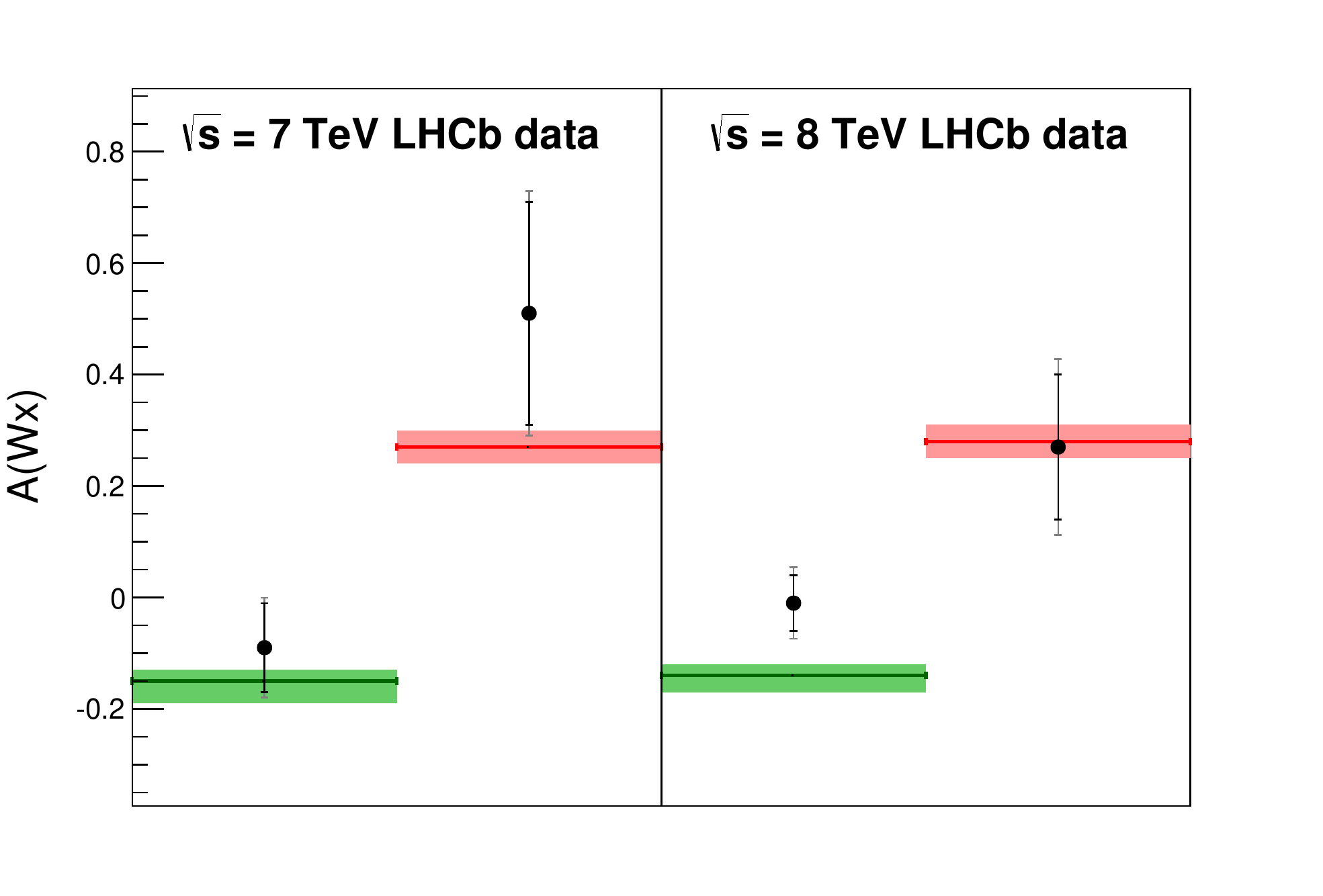}
\caption{ Comparisons of measurements performed at 7 and 8 TeV for $Wc$ (green) and $Wb$ (red) compared to theoretical predictions for (left) the ratio of $Wb$/$Wc$ to $Wj$ production and (right) $Wb$/$Wc$ charge asymmetries~\cite{Aaij:2015cha}.}
\label{fig:wbcresults}
\end{figure}

The extracted signal yields are then corrected for detector efficiency and other reconstruction effects and measurements are performed at both 7 and 8 TeV. The ratios of $Wb$ and $Wc$ production to inclusive $Wj$ production, and their charge asymmetries are shown in Figure~\ref{fig:wbcresults} and compared to the SM predictions obtained at NLO using MCFM~\cite{Campbell:2000bg} and the CT10 PDF set. A good agreement is in general observed.

\section{Electroweak $W$ and $Z$ boson production}
While the dominant mechanism for the production of $W$ and $Z$ bosons is through the annihilation of quark-anti-quark pairs, a contribution from the purely electroweak $t$-channel exchange of electroweak bosons, known as electroweak production, is also present. These amplitudes include contributions from vector-boson fusion which is of particular interest due to its similarities with Higgs boson production as well as a probe of anomalous $WWZ$ triple gauge couplings. Electroweak boson production is characterised by a boson produced in association with a pair of jets separated by a rapidity gap and with a large di-jet invariant mass. These properties can be used to distinguish it from $q\bar{q}$ annihilation and the extra phase space afforded by including jets produced in the forward region allows the signal purity to be increased by extending the rapidity gap to large values of rapidity separation, $|\Delta\eta_{jj}|$.

\begin{figure}
\includegraphics[width=0.5\textwidth]{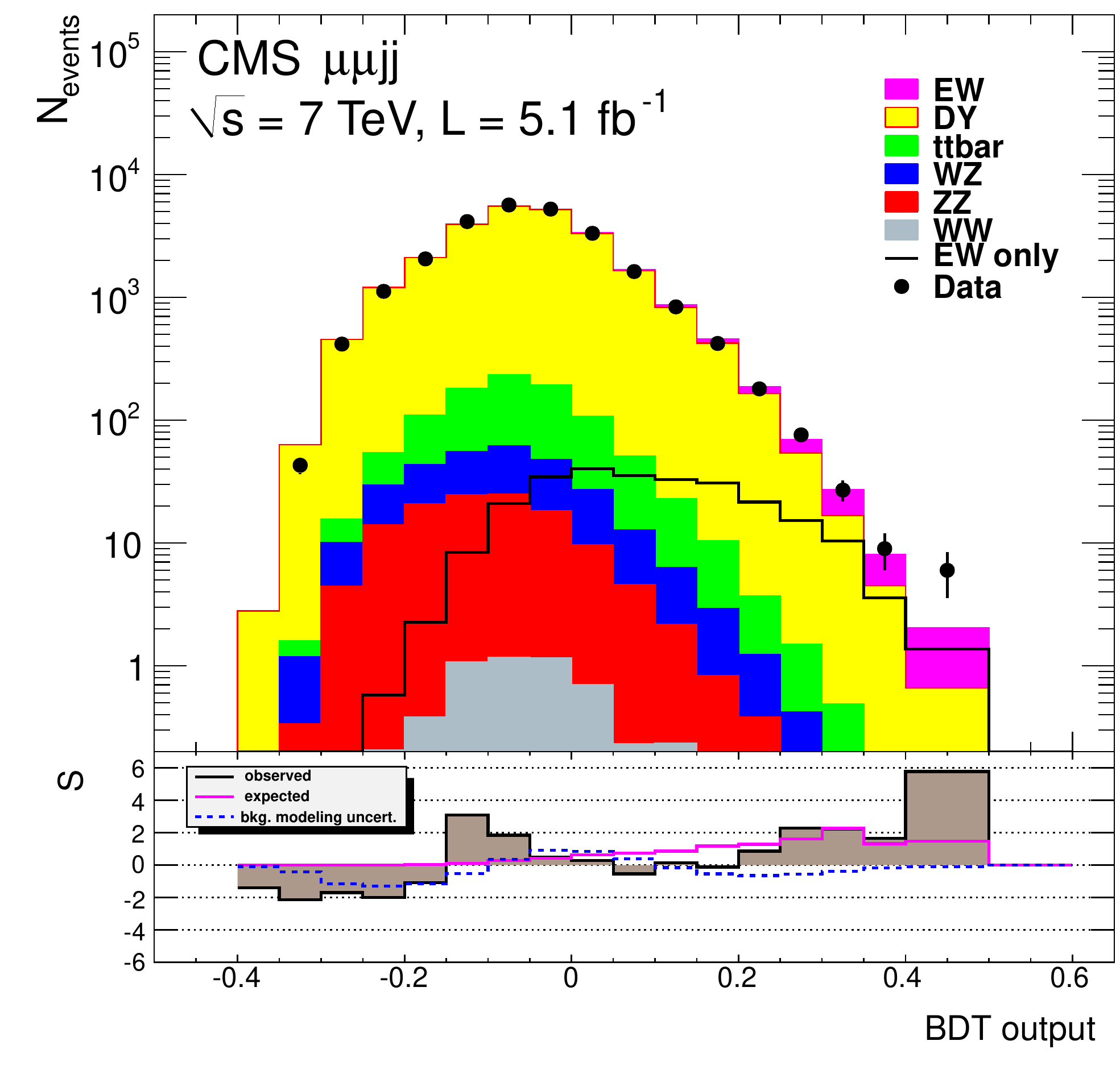}
\includegraphics[width=0.5\textwidth]{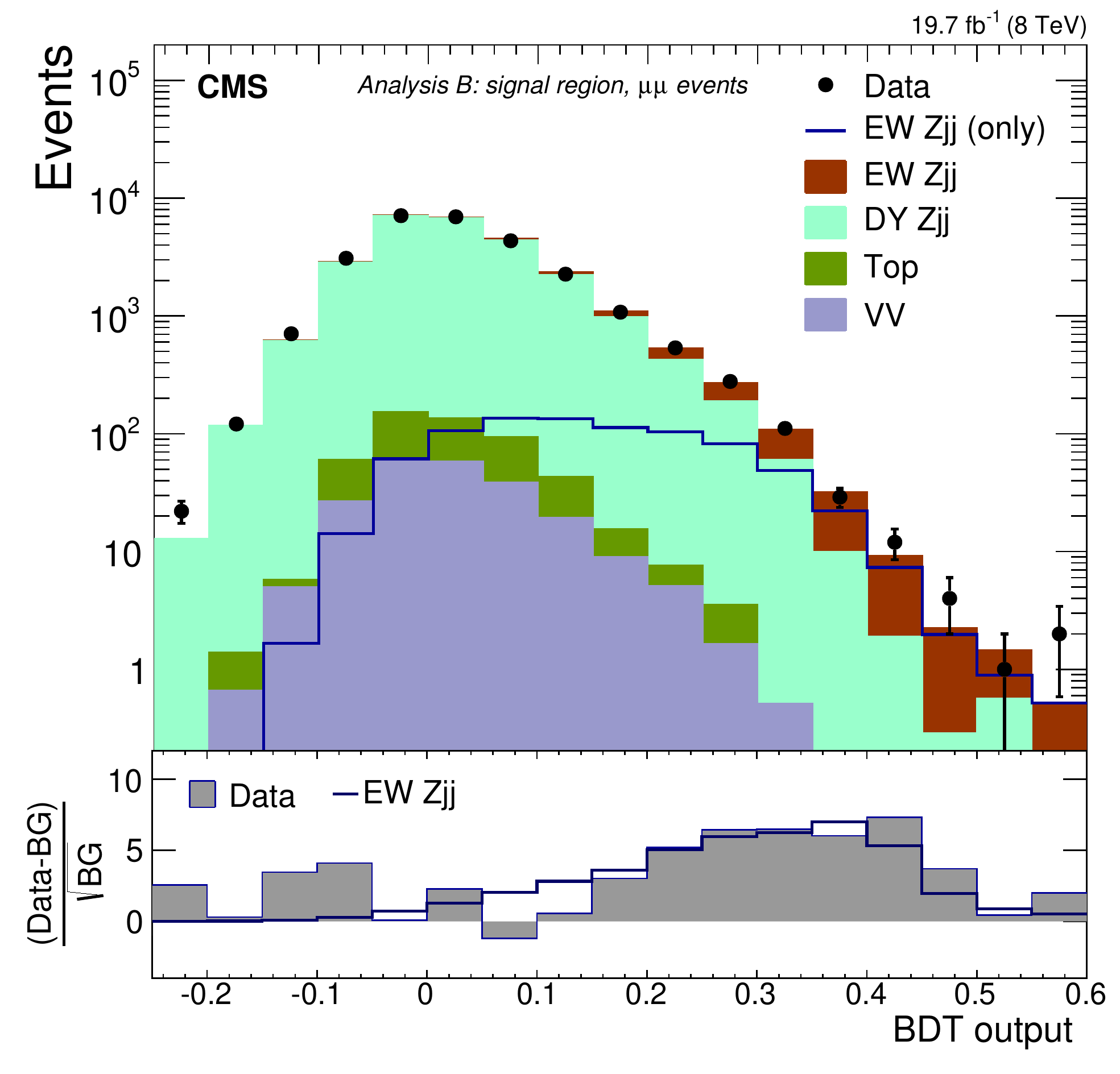}
\caption{The observed BDT responses for the dimuon channel at (left) 7~\cite{Chatrchyan:2013jya} and (right) 8 TeV~\cite{Khachatryan:2014dea} where the electroweak $Zjj$ component is visible at high BDT response values. Only one of the analysis classes is shown for the 8~TeV measurement. }
\label{fig:zjjbdts}
\end{figure}

The CMS collaboration has reported measurements of electroweak $Zjj$ production at both 7~\cite{Chatrchyan:2013jya} and 8 TeV~\cite{Khachatryan:2014dea}. $Z$ bosons are selected through both their electronic and muonic decay modes where the leptons are required to have $\pt>20$~GeV and a combined invariant mass of greater than 50~GeV. Two jets are then required to be present with transverse momenta $\pt^{j} > 25$~GeV, a di-jet invariant mass, $M_{jj}$ of greater than 120 GeV and jet pseudorapidities, $\eta_j$ of up to 4.7. Both analyses exploit multi-variate algorithms in order to separate signal from background. A BDT is trained using a number of discriminating variables, including $|\Delta\eta_{jj}|$, and the signal contribution is estimated by a template fit to the BDT response where the signal and the Drell-Yan background are free to float in the fit and the other contributions are fixed using simulation. The 8~TeV analysis makes use of three different analysis strategies, where different kinematic requirements are applied and different variables are input to the BDT. The resultant distributions for the measurements at 7 and 8~TeV for the dimuon channel are shown in Fig.~\ref{fig:zjjbdts}. with just one analysis strategy for the 8~TeV analysis shown for clarity. The cross-sections are extracted at both centre-of-mass energies with good agreement observed with SM predictions.

\begin{figure}
\includegraphics[width=0.54\textwidth]{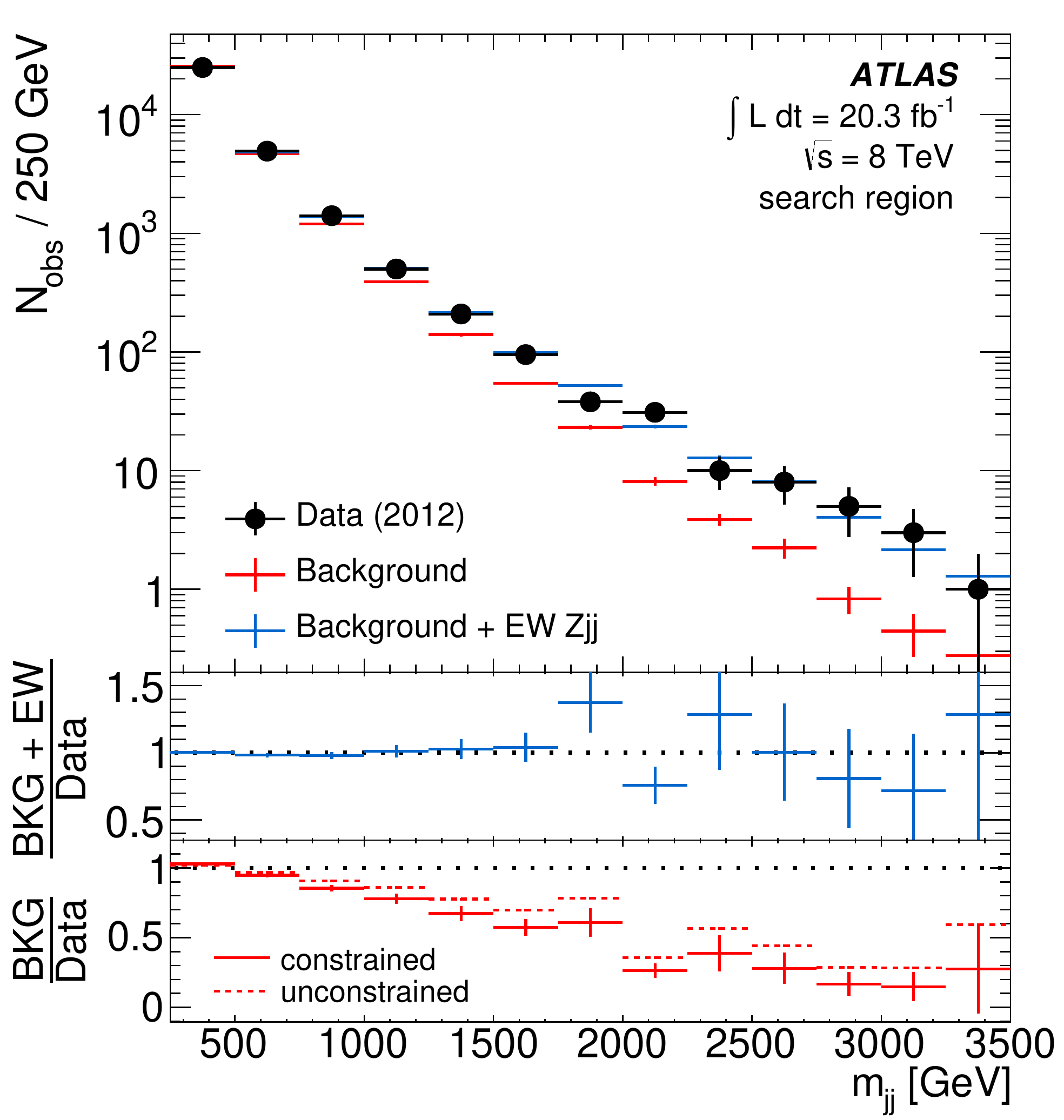}
\includegraphics[width=0.46\textwidth]{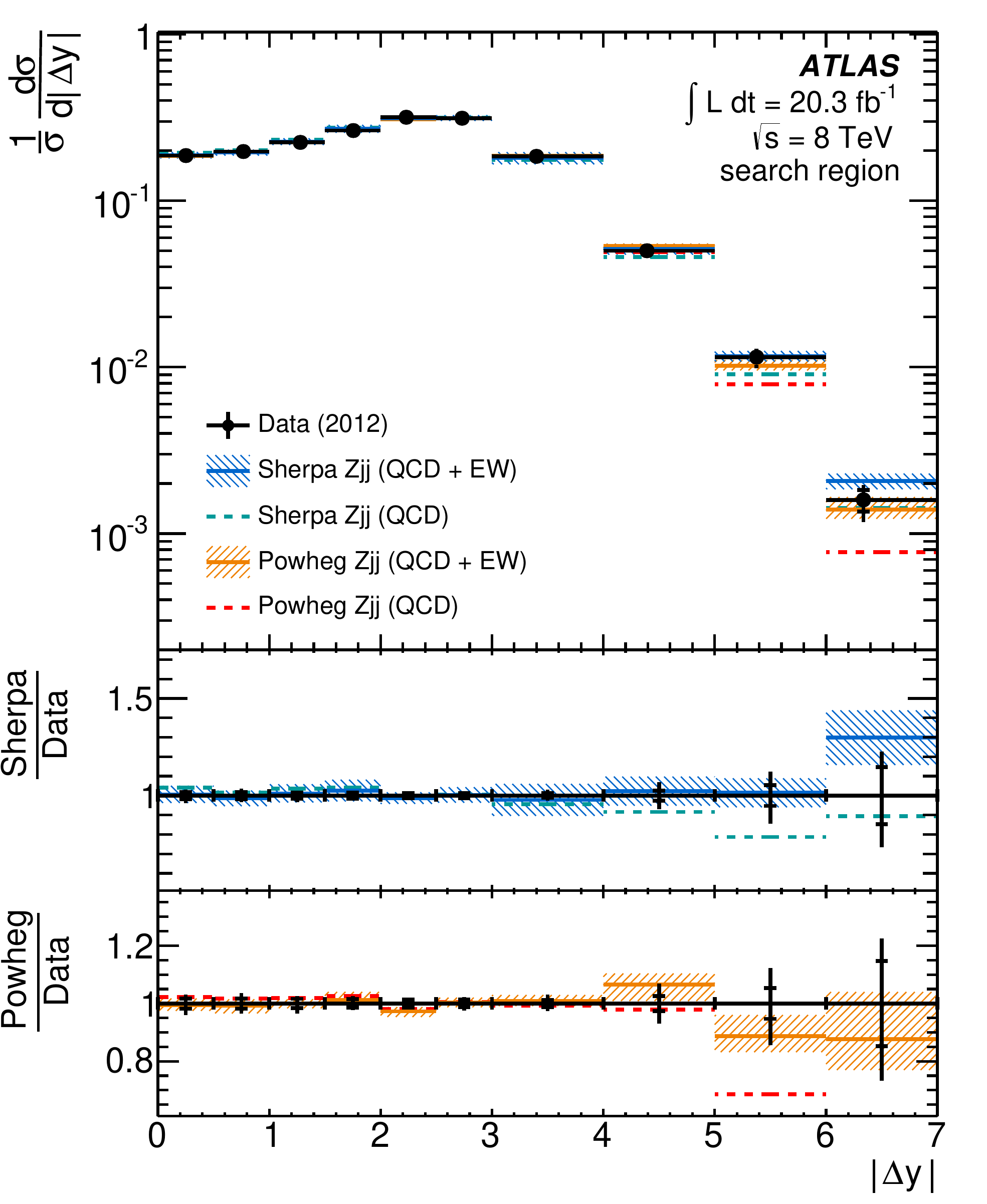}
\caption{(left) the number of observed data events compared to the expected background and signal contributions as a function of $m_{jj}$ (right) the measured cross-section as a function of $|\Delta y|$ compared to theoretical predictions~\cite{Aad:2014dta}. }
\label{fig:atlaszjj}
\end{figure}

The ATLAS collaboration has reported measurements of electroweak $Zjj$ production at 8~TeV~\cite{Aad:2014dta}.The analysis is also performed using both the electron and muon decay modes of the $Z$, with both the leptons and jets required to have a $\pt$ of greater than 25~GeV, where the jets extend up to rapidities of 4.4.  The $Zjj$ signal contribution, and subsequently the cross-section, is extracted from a fit to the dijet invariant mass in five separate fiducial regions. One of these regions, known as the ``search'' region, is chosen specifically to enhance the electroweak component by requiring a balanced $Zjj$ system in addition to a di-jet invariant mass of greater than 250~GeV and a jet rapidity gap between the two selected jets. The distributions for the search region are shown as a function of the rapidity separation and the di-jet invariant mass in Figure~\ref{fig:atlaszjj}. A good agreement is observed between the measured cross-sections and the SM predictions in the different fiducial regions, while the measurements are also used to set limits on anomalous triple gauge couplings.

\begin{figure}
\includegraphics[width=0.5\textwidth]{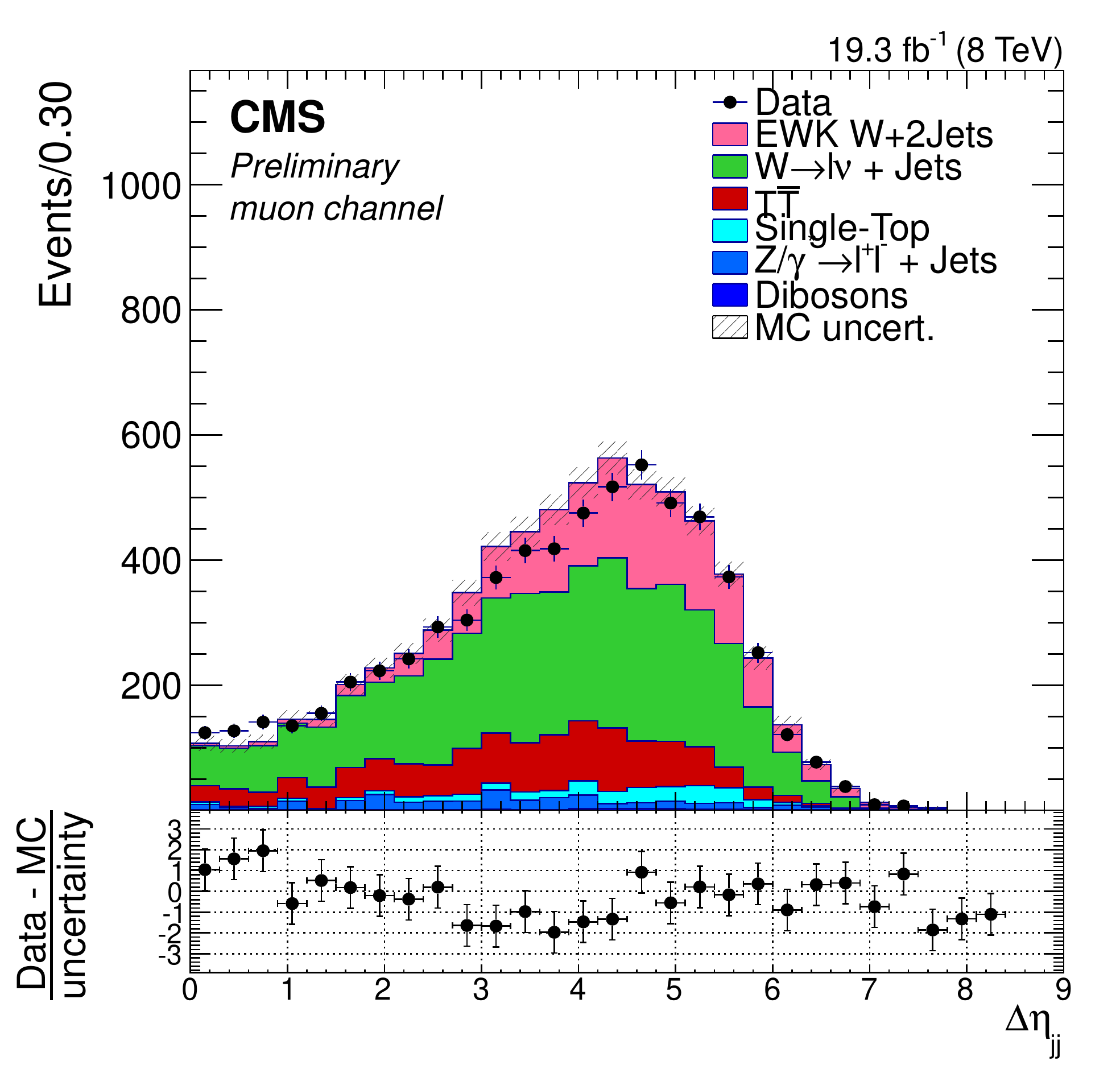}
\includegraphics[width=0.5\textwidth]{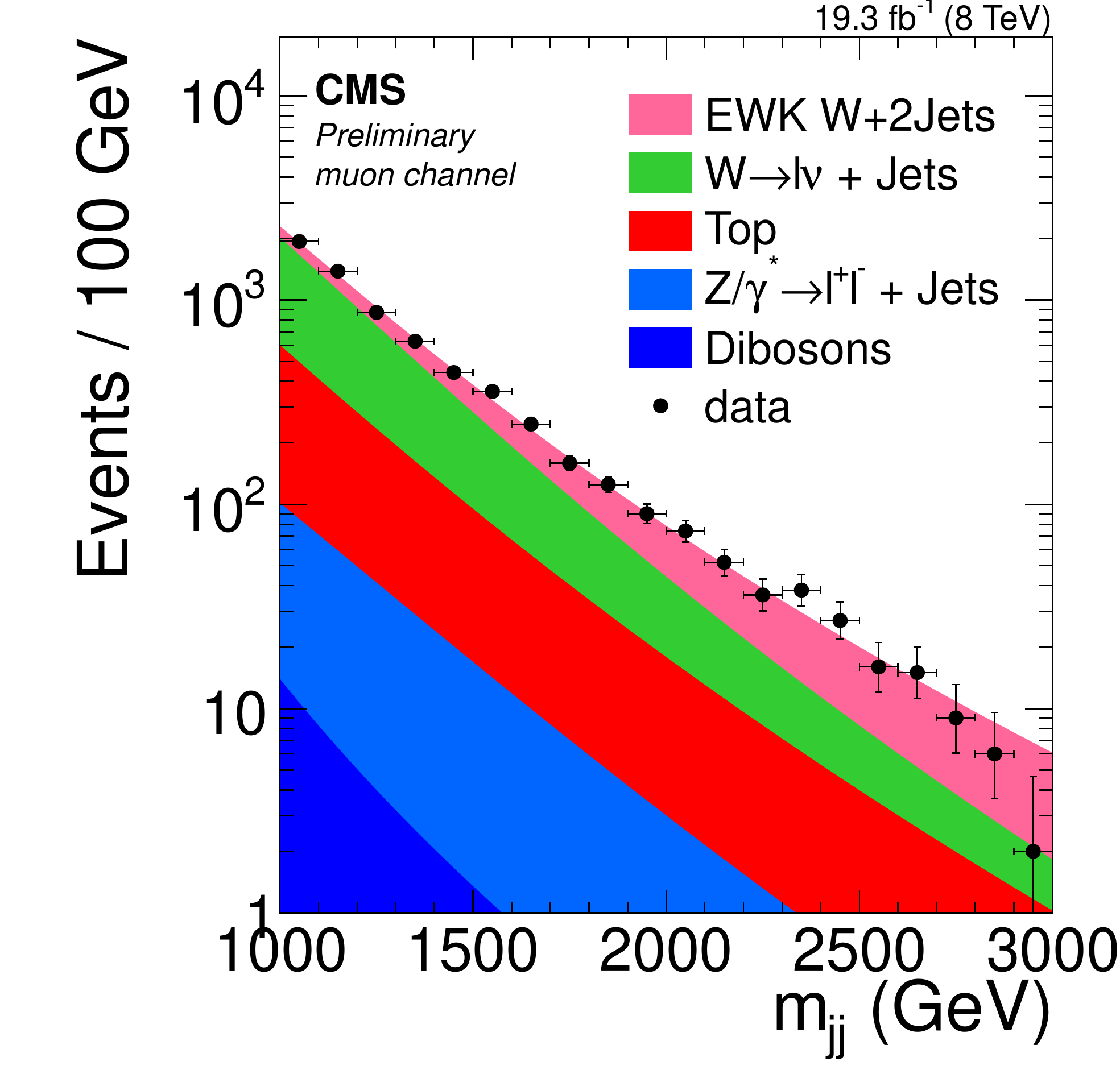}
\caption{(left) the pseudorapidity separation of the two leading jets as observed in data and compared to expectations (right) the fit used to extract the electroweak component of $Wjj$ production~\cite{CMS-PAS-SMP-13-012}.}
\label{fig:cmswjj}
\end{figure}

CMS have additionally performed a measurement of electroweak $Wjj$ production~\cite{CMS-PAS-SMP-13-012} in both the muon and electron decay mode where slightly different kinematic requirements are performed compared to the $Zjj$ measurement. Muons (electrons) are required to have $\pt$ greater than 25 (30)~GeV and $|\eta|<$ 2.1(2.5). Two jets are also required with a pseudorapidity of less than 4.7 where the leading jet has a $\pt$ of greater than 60~GeV and the sub-leading jet has a $\pt$ of greater than 50~GeV. The transverse missing energy is required to be greater than 25(30) GeV for the muon(electron) channel. The cross-section is extracted from a fit to the $M_{jj}$ spectrum, shown in Figure~\ref{fig:cmswjj} along with the $\Delta\eta_{jj}$ distribution observed in data and compared to expectations. A good agreement is observed with the SM predictions.

\section{Forward-Backward Asymmetry and extraction of $\sin^2\theta_W$ in $Z\to\ell\ell$ events}
\label{sec:sin2tw}
The annihilation process $q\bar{q}\to\ell^+\ell^-$ exhibits a forward-backward asymmetry, $A_{\rm FB}$, due to the presence of both vector and axial-vector amplitudes. The asymmetry is observed when considering the distribution of the polar angle of the positive lepton, measured with respect to the quark direction in the $q\bar{q}$ rest frame. It shows a strong dependence on the dilepton invariant mass near the $Z$ resonance. As the LHC is a symmetric $pp$ collider, the quark direction is not known and the forward direction is alternatively defined with respect to the $z$ component of the $Z$ boson momentum. This results in a dilution of the asymmetry in cases where the $Z$ boson does not follow the direction of the initial quark. As the valence quarks are more prominent in the forward region, a consequence of the higher $x$-region probed, the parton level asymmetry is more pronounced in this region.

The ATLAS, CMS and LHCb experiments have performed measurements of $A_{\rm FB}$ as a function of dilepton invariant mass using Run-I data.  ATLAS has performed the measurement at 7~TeV~\cite{Aad:2015uau} while CMS has performed the measurement at both 7 and 8~TeV~\cite{Chatrchyan:2012dc, CMS-PAS-SMP-14-004} where the 7~TeV result does not include forward electrons and consequently is not discussed here. The measurements have been performed in both the dimuon and dielectron channels where the final state leptons are required to be in the central region. Additionally, in order to access the more sensitive forward region, the dielectron channel is extended by requiring that one of the electrons be reconstructed in the central region, and the second be reconstructed in the forward calorimeters. LHCb has performed the measurement in the dimuon channel at 7 and 8 TeV~\cite{Aaij:2015lka}, where both muons are reconstructed in the forward region. A summary of the different channels and kinematic regions explored by the three experiments is shown in Table~\ref{tab:afb}.

\begin{table}
{\footnotesize
\caption{A summary of the different leptonic decay modes and kinematic ranges used by the LHCb, CMS and ATLAS collaborations to study the forward-backward asymmetry. ATLAS and CMS both explore ``central-forward'' regions, where one of the leptons is required to be central and the other to be in the forward region.}
\label{tab:afb}
\begin{tabular}{c c c c c}
Exp. & Channel & $M_{\ell\ell}$ & $p_{\rm T}^\ell$ & $\eta^\ell$ \\
\hline
LHCb & dimuon & $60 - 160$~GeV & $ > 20$~GeV & $2 < \eta < 4.5$ \\
CMS & dimuon & $40 - 2000$~GeV & $> 20$~GeV  & $|\eta| < 2.4$ \\
CMS & dielectron & $40 - 2000$~GeV & $>20$~GeV  & $|\eta| < 2.4$ \\
CMS & central-forward electron & $40 - 300$~GeV & $> 30,20$~GeV  & $|\eta| < 2.4$, $3.0<\eta<5$ \\
ATLAS & dimuon & $40 - 2000$~GeV &$ > 25$~GeV  & $|\eta| < 2.4$ \\ 
ATLAS & dielectron & $40 - 1000$~GeV &$> 25$~GeV  & $|\eta| < 2.47$ \\
ATLAS & central-forward electron & $40 - 250$~GeV & $> 25$~GeV  & $|\eta| < 2.47$, $2.5<\eta<4.9$\\
\end{tabular}
}
\end{table}

\begin{figure}
\includegraphics[width=0.32\textwidth]{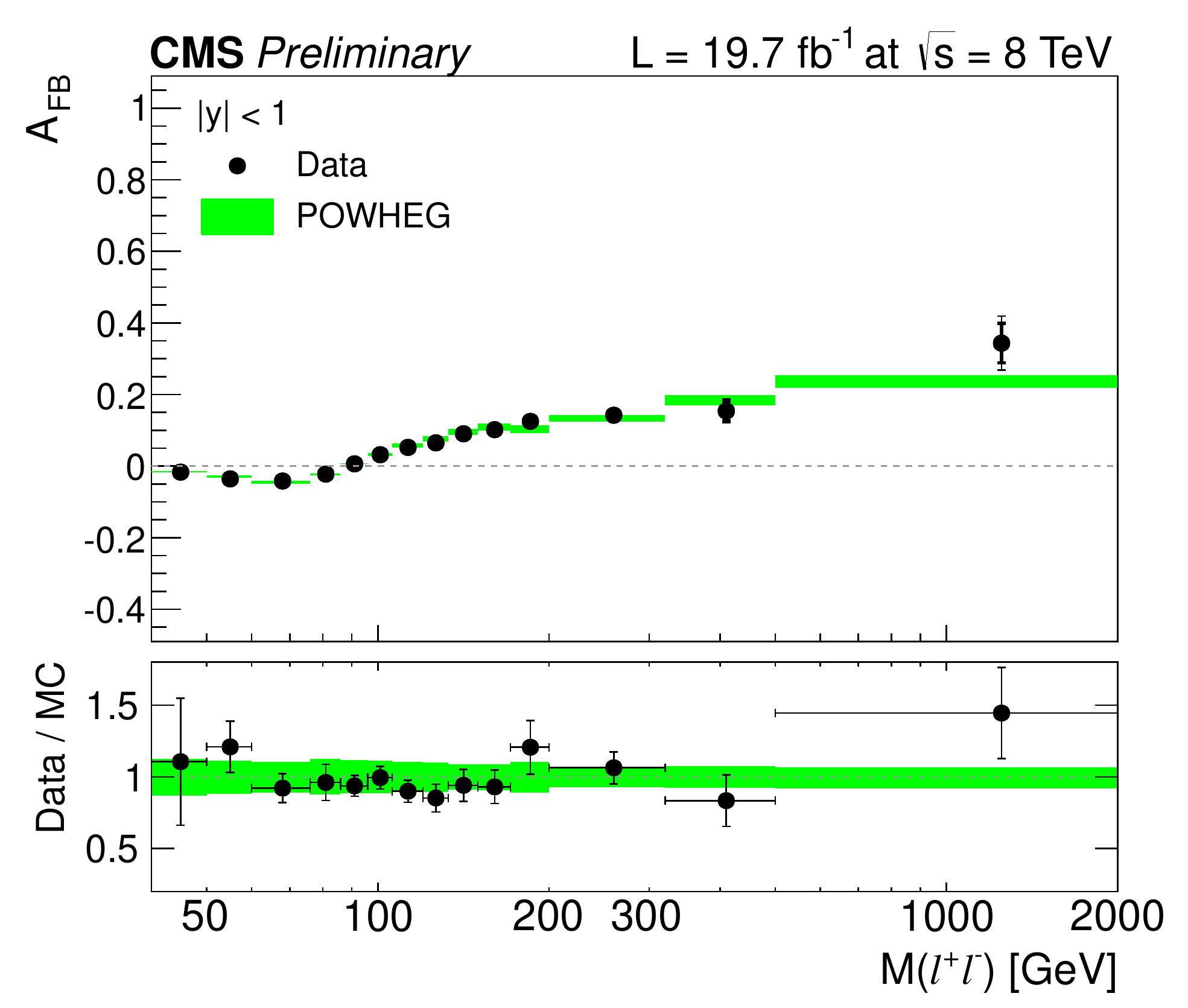}
\includegraphics[width=0.32\textwidth]{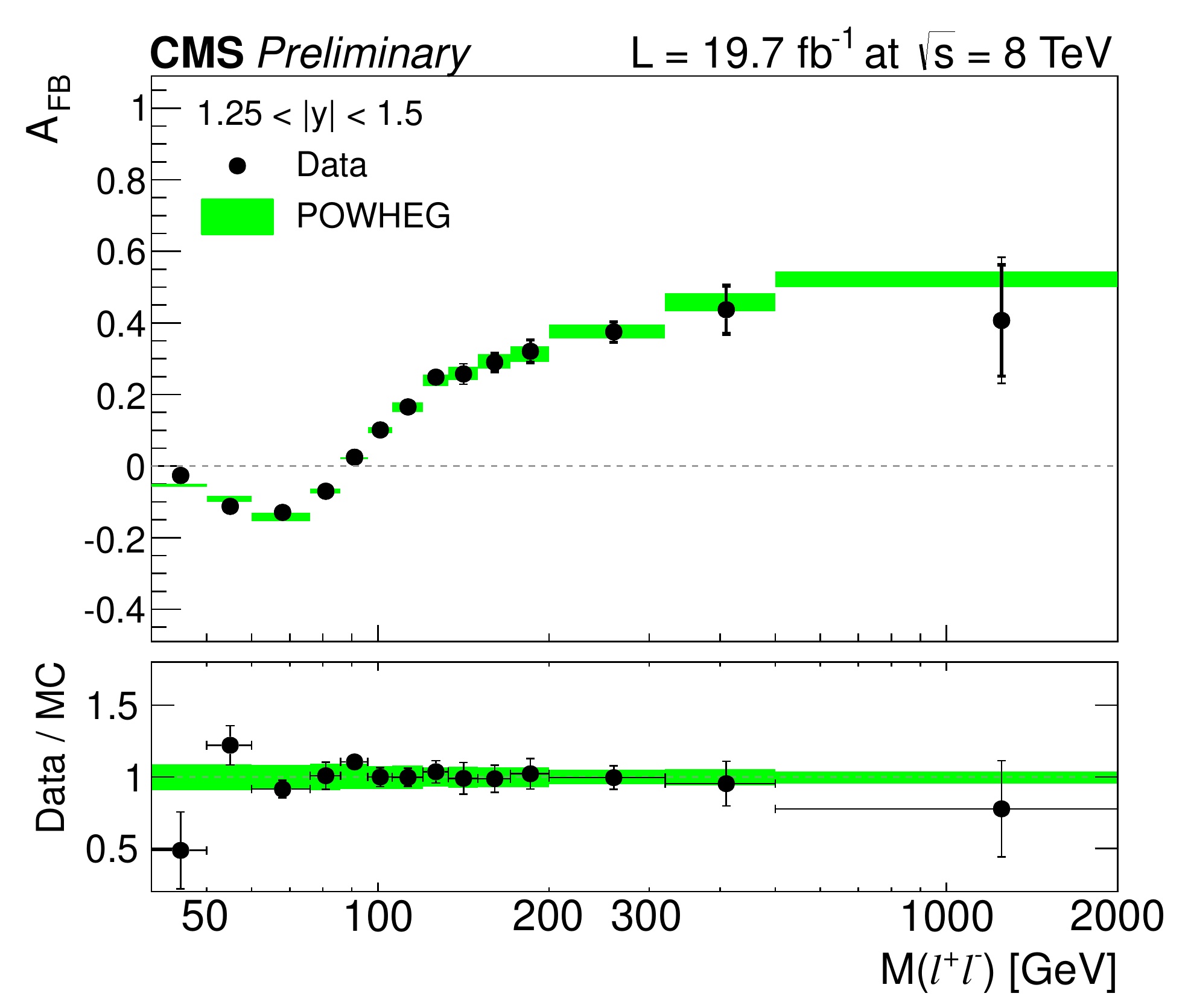}
\includegraphics[width=0.32\textwidth]{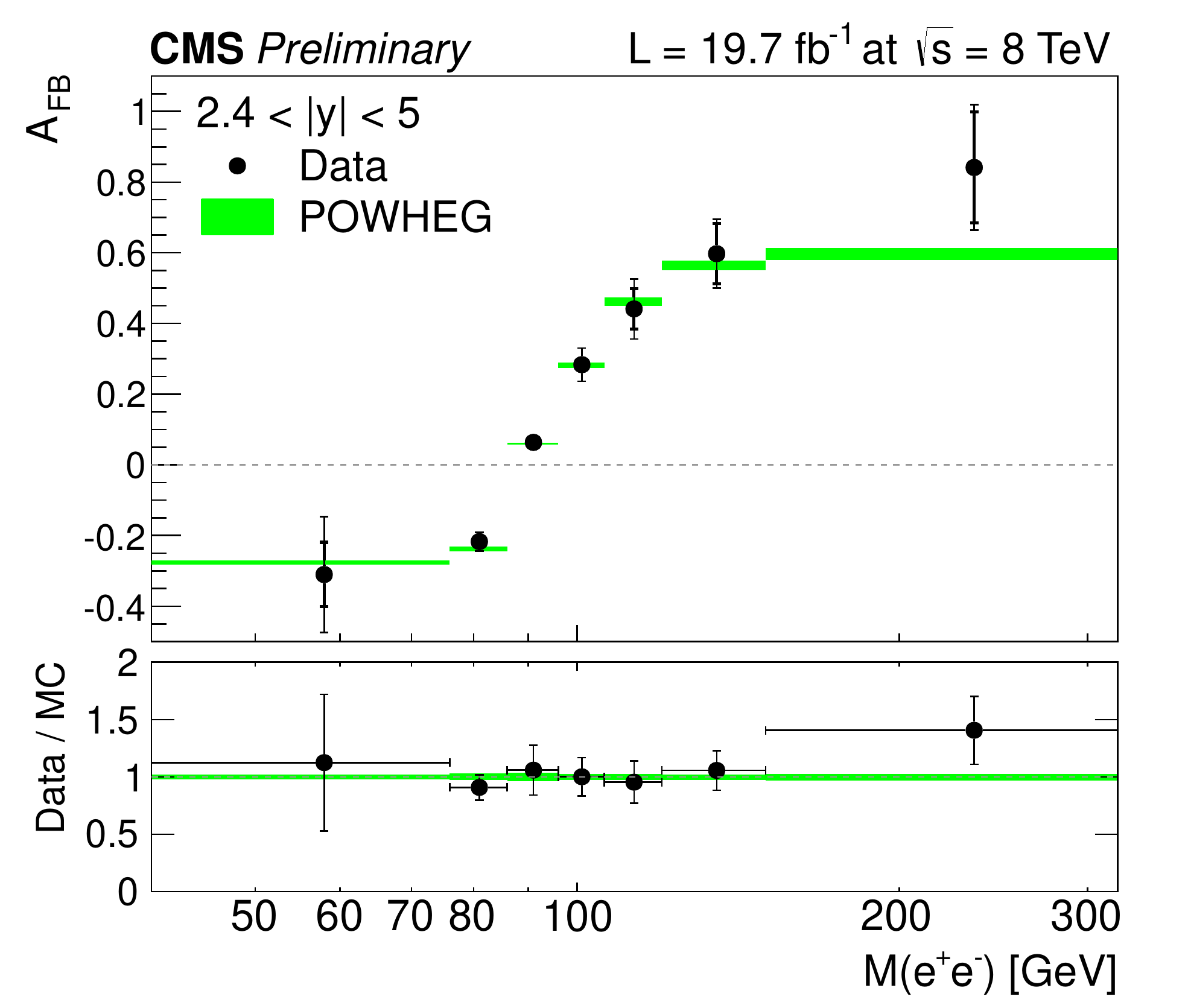}
\caption{The forward backward asymmetry as measured by the CMS collaboration in three different rapidity bins. The muon and electron channels are combined in the central region, while the measurement in the forward region is only performed in the electron channel~\cite{CMS-PAS-SMP-14-004}.}
\label{fig:afb_cms}
\end{figure}

\begin{figure}
\includegraphics[width=0.44\textwidth]{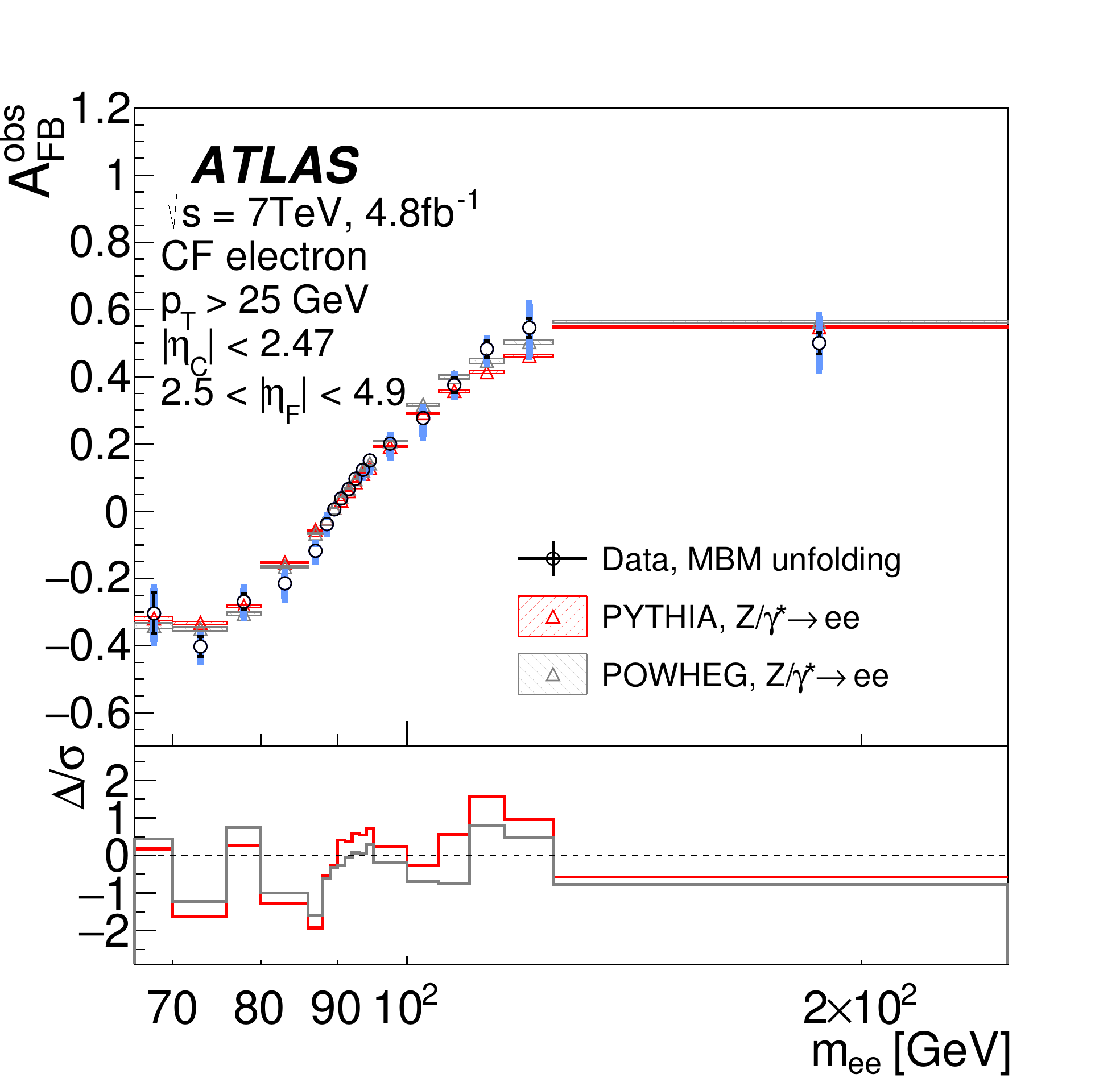}
\includegraphics[width=0.56\textwidth]{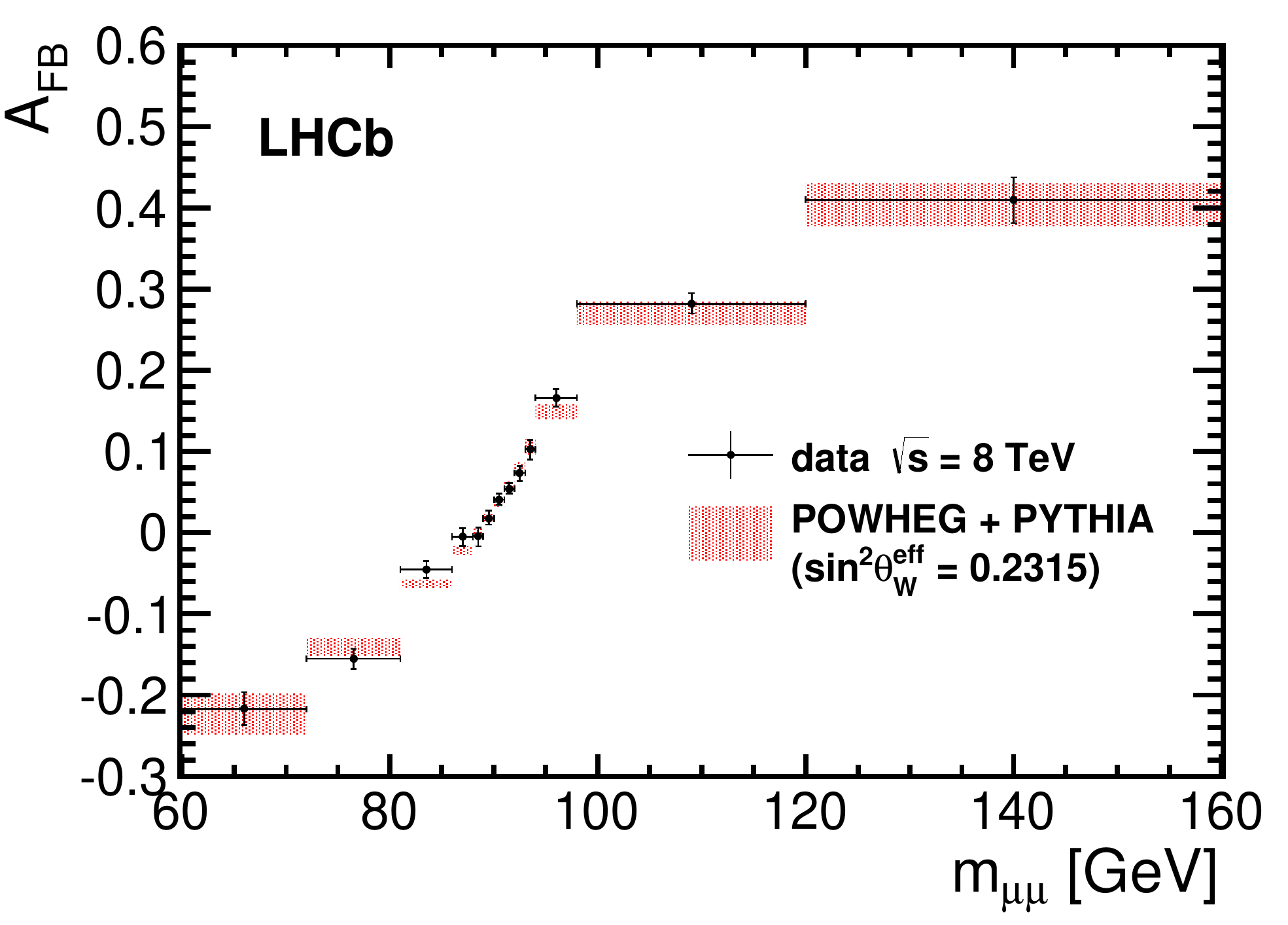}
\caption{The forward backward asymmetry measured by (left) ATLAS in the central-forward region~\cite{Aad:2015uau} and (right) LHCb~\cite{Aaij:2015lka}. The ATLAS measurement is shown before correction for dilution and is compared to predictions generated using both \textsc{Pythia}  and \textsc{Powheg} while the LHCb measurement is compared to \textsc{Powheg}.}
\label{fig:afb_atlaslhcb}
\end{figure}

The distributions obtained from data are corrected for background contributions, as well as unfolded for detector effects to obtain the true asymmetry distributions. In the case of the ATLAS measurement, the distribution is additionally corrected for the effects of dilution using \textsc{Pythia}~\cite{Sjostrand:2006za} simulation. The ATLAS and LHCb result is performed as a function of dilepton invariant mass, while CMS performs the measurement double-differentially as a function of invariant mass and rapidity. The forward-backward asymmetry as measured by CMS is shown for three different rapidity bins in Figure~\ref{fig:afb_cms} where the increasing asymmetry is evident moving from the central to the forward region. The results for LHCb and ATLAS are shown in Figure~\ref{fig:afb_atlaslhcb}, where the central-forward region is shown for ATLAS.

\subsection{Extraction of $\sin^2\theta_W^{\rm eff}$}
\begin{figure}
\begin{center}
\includegraphics[width=0.6\textwidth]{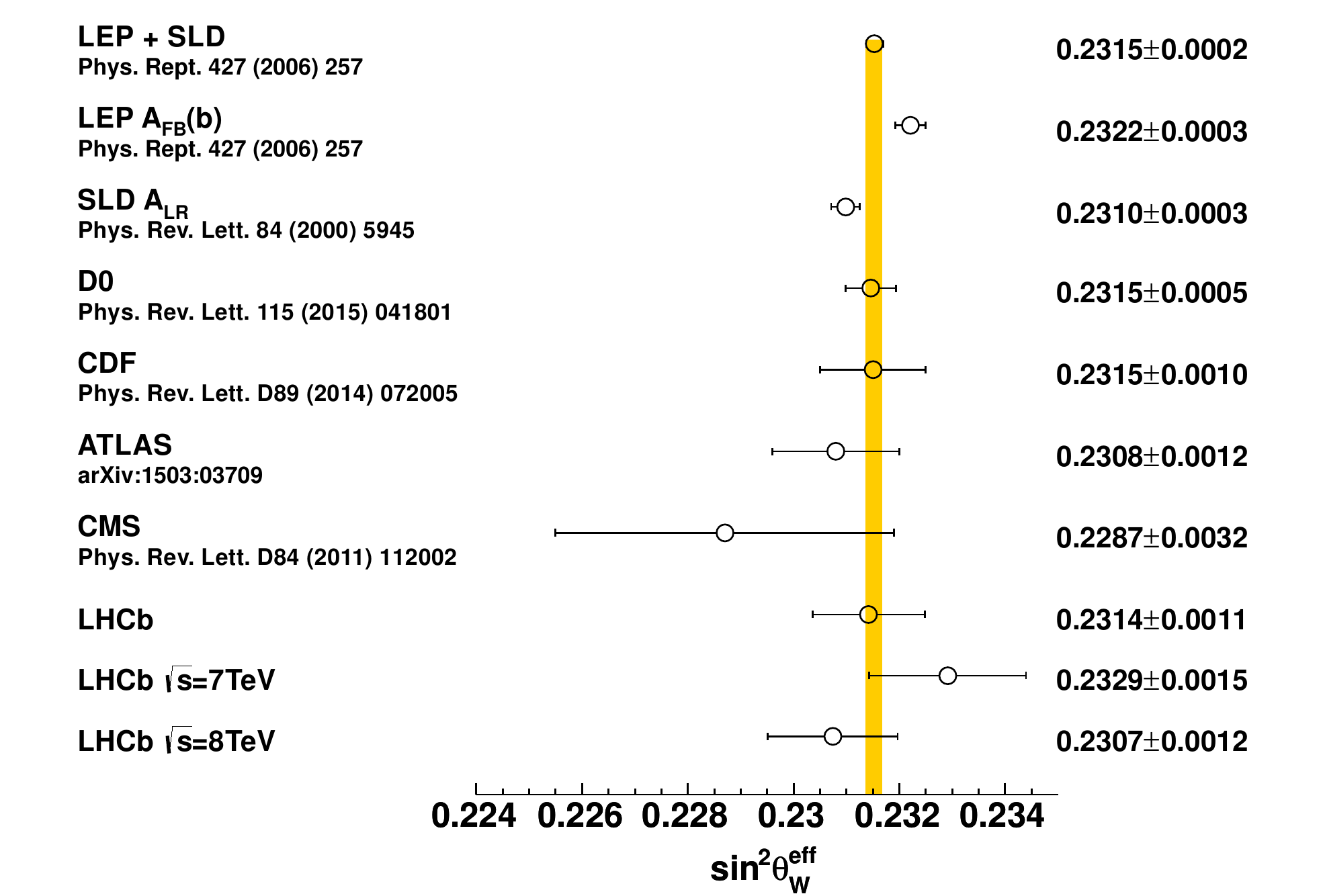}
\end{center}
\caption{Comparison of measurements of $\sin^2\theta_W$ as performed by ATLAS, LHCb and a number of different experiments~\cite{Aad:2015uau}.}
\label{fig:afb_results}
\end{figure}

As $A_{\rm FB}$ is sensitive to the effective weak mixing angle, $\sin^2\theta_W^{\rm eff}$, both ATLAS and LHCb have extended the measurement to extract the value of $\sin^2\theta_W^{\rm eff}$ through a template fit to the measured asymmetry. Predictions are obtained using a range of values of $\sin^2\theta_W^{\rm eff}$ and a $\chi^2$ minimisation is performed in order to determine the best fit value. The templates are generated using \textsc{Powheg}~\cite{Alioli:2008gx} for LHCb and \textsc{Pythia} for ATLAS. A comparison of the results, in addition to other measurements is shown in Figure~\ref{fig:afb_results}. The dominant source of systematic uncertainty for both measurements is due to the description of the PDFs, which is lower for LHCb due to the higher values of boson rapidity probed than the central or forward-central measurements performed at ATLAS. The PDFs also represent the largest uncertainty on the ATLAS measurement, while the LHCb measurement is limited by the available statistics.

\section{Conclusion}
Measurements have been presented of electroweak production in the forward region, where the final state involves forward leptons and/or jets in the final state. The LHCb detector, situated in the forward region, has performed measurements of both inclusive and associated $W$ and $Z$ boson production. The ATLAS and CMS collaboration have both performed measurements including forward jets for the the study of the electroweak production of electroweak bosons, while ATLAS has additionally performed measurements of $W$ and $Z$ production in association with forward jets. All three experiments have performed measurements of the forward-backward asymmetry in $Z\to\mu\mu$ events, where LHCb selects dilepton pairs in the forward region, and ATLAS and CMS exploit both the central, and ``central-forward'' regions, where calorimeters are used to select dielectron events with one electron in the forward region.




\setboolean{inbibliography}{true}
\bibliographystyle{LHCb}%
\bibliography{lhcp}%

\end{document}